\documentclass[sn-mathphys-num]{sn-jnl}% Math and Physical Sciences Numbered Reference Style
\usepackage{graphicx}%
\usepackage{multirow}%
\usepackage{amsmath,amssymb,amsfonts}%
\usepackage{amsthm}%
\usepackage{mathrsfs}%
\usepackage[title]{appendix}%
\usepackage{xcolor}%
\usepackage{textcomp}%
\usepackage{manyfoot}%
\usepackage{booktabs}%
\usepackage{siunitx}%
\usepackage{algorithm}%
\usepackage{algorithmicx}%
\usepackage{algpseudocode}%
\usepackage{listings}%
\usepackage{braket}%
\usepackage{lipsum}%
\usepackage{pgfplots}
\usepgfplotslibrary{groupplots}
\usepackage{pgfplotstable}
\usepackage{subcaption}
\usepackage{filecontents}
\usepackage{graphicx}
\usepackage{caption}
\usepackage{placeins}
\pgfplotsset{compat=1.18}
\usepgfplotslibrary{statistics}
\usepackage{array}    % Required for fancy column types

\usepackage{comment}

\usepackage[commandnameprefix=ifneeded]{changes}

\definechangesauthor[color=red]{EA}
\definechangesauthor[color=blue]{GB}
\definechangesauthor[color=brown]{CC}
\theoremstyle{thmstyleone}%
\theoremstyle{thmstyletwo}%

\theoremstyle{thmstylethree}%

\begin{document}

\title[QUBO-based training for VQAs on Quantum Annealers]{QUBO-based training for VQAs on Quantum Annealers}

\author*[1]{\fnm{Ernesto} \sur{Acosta}}\email{eacostam@correo.ugr.es}

\author[2]{\fnm{Guillermo} \sur{Botella}}\email{gbotella@ucm.es}

\author[1]{\fnm{Carlos} \sur{Cano}}\email{carloscano@ugr.es}

\affil[1]{\orgdiv{Dpt. Computer Science and AI}, \orgname{University of Granada}, \orgaddress{\city{Granada}, \postcode{18071}, \state{Andalucía}, \country{Spain}}}

\affil[2]{\orgdiv{Computer Architecture and Automation Department}, \orgname{Complutense University of Madrid}, \orgaddress{\city{Madrid}, \postcode{28040}, \state{Madrid}, \country{Spain}}}

\abstract{

Quantum annealers provide an effective framework for solving large-scale combinatorial optimization problems. This work presents a novel methodology for training Variational Quantum Algorithms (VQAs) by reformulating the parameter optimization task as a Quadratic Unconstrained Binary Optimization (QUBO) problem. Unlike traditional gradient-based methods, our approach directly leverages the Hamiltonian of the chosen VQA ansatz and employs an adaptive, metaheuristic optimization scheme. 
This optimization strategy provides a rich set of configurable parameters which enables the adaptation to specific problem characteristics and available computational resources. The proposed framework is generalizable to arbitrary Hamiltonians and integrates a recursive refinement strategy to progressively approximate high-quality solutions. 

Experimental evaluations demonstrate the feasibility of the method and its ability to significantly reduce computational overhead compared to classical and evolutionary optimizers, while achieving comparable or superior solution quality.

These findings suggest that quantum annealers can serve as a scalable alternative to classical optimizers for VQA training, particularly in scenarios affected by barren plateaus and noisy gradient estimates, and open new possibilities for hybrid quantum gate - quantum annealing - classical optimization models in near-term quantum computing.
}


\maketitle

\section{Introduction}\label{sec1}

Training parameterized quantum circuits using Variational Quantum Algorithms (VQAs) constitutes a fundamental strategy to exploit near-term quantum devices in applications such as optimization, quantum chemistry, and machine learning~\cite{cerezo2024vqa_review}. These hybrid quantum-classical algorithms iteratively update circuit parameters to minimize a cost function, seeking global or high-quality local minima. However, optimization is hindered by the non-convexity of quantum cost landscapes, often shaped by entanglement, hardware constraints, and noise~\cite{franca2020limitations}. The choice of a classical optimizer strongly affects convergence and performance~\cite{bonetmonroig2021performance}. A critical limitation arises from barren plateaus, where gradients vanish exponentially with system size, preventing parameter updates~\cite{mcclean2018barren}. Moreover, the noise intrinsic to quantum measurement adversely affects gradient estimates~\cite{wang2021noise}. To address these issues, various techniques have emerged, including layer wise training~\cite{skolik2021layerwise}, adaptive learning rates~\cite{liu2018differentiable}, quantum natural gradient methods~\cite{stokes2020quantum}, and problem tailored ansätze~\cite{Schuld2014}, alongside quantum assisted optimization and physics informed heuristics~\cite{Schuld2019}.

In this work, we introduce a novel optimization framework that employs a Quantum Annealer to determine optimal rotation angles for VQA ansätze, thereby alleviating limitations of classical training methods such as barren plateaus and slow convergence. Our method utilizes the intrinsic properties of quantum annealing and formulates the training task as a recursive QUBO-based approximation, enabling high-quality solutions with reduced computational overhead. 

This paper is organized as follows. Section~\ref{relatedwork} reviews related efforts involving VQA training and QUBO applications. Section~\ref{methods} details the proposed model, while Section~\ref{experiment} presents the experimental setup and comparative results against classical and evolutionary optimizers. This section includes an analysis of the model's scalability in terms of dataset size and recursive depth, and a discussion on the model’s behavior under noise. Finally, Section~\ref{conclusions} summarizes the main findings and implications of this work.

\section{Related Work}\label{relatedwork}

Several recent efforts have sought to improve the training of VQAs by exploring hybrid classical-quantum models, alternative hardware platforms, and reformulations of the optimization process.

Classical optimization remains foundational for VQA training. Gradient-based methods ~\cite{fletcher1970,shanno1970,lockwood2022empirical} are commonly applied to smooth cost functions using Hessian approximations, while the Frank–Wolfe algorithm~\cite{FrankWolfe1956} provides a projection-free method for constrained problems. While mature and well-characterized, these methods face challenges in the quantum domain, where non-convex cost landscapes and barren plateaus hinder effective training increasing interest in alternative approaches, including evolutionary and quantum assisted techniques.

Evolutionary Algorithms (EAs) have also been proposed for VQA Training. The EVOVAQ framework~\cite{ACAMPORA2024101756}, developed in Python and integrated with Qiskit,  offers resource-efficient VQA optimization based on EAs and suited to noisy intermediate scale quantum (NISQ) devices. By avoiding gradient-based limitations, EVOVAQ enhances trainability and flexibility in navigating complex parameter landscapes, contributing to more robust hybrid classical-quantum algorithm design.
   
On another side, Quadratic Unconstrained Binary Optimization (QUBO) Formulations have been proposed as well as a training mechanism in traditional Machine Learning. Previous work by Date et al.~\cite{Date2021} reformulated classical machine learning models, including linear regression, support vector machines (SVMs), and balanced k-means clustering into QUBO problems suitable for adiabatic quantum computing (AQC). These reformulations yielded comparable or improved computational performance relative to traditional approaches. However, the current methods do not yet extend to deep learning architectures, leaving a gap in applying QUBO-based training to modern, non linear models.

At the same time, AQC has been proposed as training infrastructure for Neural Networks~\cite{10.3389/frai.2024.1368569}, particularly suited for models with discrete weights, such as Binary Neural Networks (BNNs), the approach leverages quantum tunneling to explore combinatorial parameter spaces. The binary connectivity formulation aligns with current quantum hardware limitations and was validated through a cosine approximation task, though broader benchmarks are needed for evaluating scalability and generalization.

Pramanik et al.~\cite{pramanik2024training} introduced iTrust, a method employing opto-electronic oscillator-based coherent Ising machines (CIMs) with modified transfer functions for trust region optimization under box constraints. Enhancements such as non symmetric coupling, noise modulation, and convex projections were analytically shown to improve convergence. Although CIMs are inspired by quantum mechanics, they are not full quantum devices, and the study does not include a practical implementation, calling for further assessment of feasibility for real world deployment.

Finally, in a previous work, we proposed a hybrid Quantum Machine Learning (QML) model that combines classical, gate-based, and AQC to address training challenges in VQAs~\cite{acosta2024adiabatictrainingvariationalquantum, QTML2024}. 

This was a first attempt to propose a QUBO-based formulation to address this problem via quantum annealing. Preliminary results, evaluated on a small benchmark dataset, showed comparable accuracy to classical gradient-based methods, though the scope was limited to binary-encoded problems and fixed circuit structures. While these results were promising, improvements in execution time remained speculative due to hardware constraints and problem size limitations. 

In this work, we build upon our previous approach by incorporating a recursive adiabatic search strategy that mitigates the need for full-resolution one-shot training. We also describe how to generalize the QUBO formulation to represent broader classes of VQAs, propose symbolic optimizations of the quantum operator representation to enhance scalability and perform and exhaustive experimentation on three classical datasets to study the impact of the parameters on the results. 

\section{Proposed Approach}\label{methods}

The proposed approach consists of: 1) formulating the optimization task of training a variational circuit as a QUBO expression (see Sec. \ref{sec:qubo_formulation}), and 2) proposing a recursive adiabatic search using a quantum annealer to find a solution to the QUBO expression (see Sec. \ref{sec:optimization}).

\subsection{QUBO Formulation for VQA optimization}\label{sec:qubo_formulation}

\subsubsection{QUBO Formulation via State Vector Comparison}\label{secformulation}

Given a dataset $S=\{(\mathbf{x}_i, y_i) \}^{r-1}_{i=0}$ where $\mathbf{x}_i \in \mathcal{X}$ are the $r$ training instances and $y_i \in \mathcal{Y}$ are their corresponding labels, the training of VQAs typically involves minimizing the discrepancy between the expected outputs ($y_i$) and the actual outputs ($\hat{y_i}$) produced by a parameterized quantum circuit. A widely used cost function is the Mean Squared Error (MSE) which defines such difference across all samples in a dataset as:

\begin{equation}
\text{MSE} = \frac{1}{r} \sum_{i=0}^{r-1} \left( y_i - \hat{y}_i \right)^2
\label{eq:mse}
\end{equation}

In our model, both the expected and observed outputs are represented as quantum state vectors in the Hilbert space of the system.
Let $|\psi\rangle$ denote the expected state vector, derived from the label in the dataset, and $|\phi\rangle$ the state vector obtained at the output of the variational circuit. The Euclidean distance between the two vectors is used as a measure of error:
\begin{equation}
d(|\psi\rangle, |\phi\rangle) = \left\| |\psi\rangle - |\phi\rangle \right\|
\label{eq:euclidean_norm}
\end{equation}

Given the expansion of each state vector in the computational basis:
\begin{align}
|\psi\rangle &= (\psi_0, \psi_1, \dots, \psi_{2^q - 1}) \\
|\phi\rangle &= (\phi_0, \phi_1, \dots, \phi_{2^q - 1})
\end{align}
where $q$ is the number of qubits, the Euclidean distance becomes:
\begin{equation}
d(|\psi\rangle, |\phi\rangle)^2 = \sum_{j=0}^{2^q - 1} |\psi_j - \phi_j|^2
\label{eq:euclidean_squared}
\end{equation}

Since QUBO formulations operate over real valued quadratic expressions, we simplify the expression assuming real valued amplitudes (or that we use real projections of the state vectors), yielding:
\begin{equation}
d^2(|\psi\rangle, |\phi\rangle) = \sum_{j=0}^{2^q - 1} (\psi_j - \phi_j)^2
\label{eq:real_amplitudes_distance}
\end{equation}

For a dataset composed of $r$ training samples, the total cost over all records based on the MSE as defined in Equation \ref{eq:mse} is:
\begin{equation}
\mathrm{QUBO}_d = \frac{1}{r} \sum_{i=0}^{r - 1} \sum_{j=0}^{2^q - 1} \left( \psi_{ij} - \phi_{ij} \right)^2
\label{eq:qubo_total_loss}
\end{equation}
where $\psi_{ij}$ is the $j$-th component of the expected state vector for the $i$-th training record, and $\phi_{ij}$ is the corresponding component of the state vector generated by the variational circuit.

The output state $|\phi\rangle$ is obtained by applying a parameterized unitary operator $Q$ (defined by the ansatz) to the input state:
\begin{equation}
|\phi\rangle = Q |\psi\rangle = \sum_{j=0}^{2^q - 1} \left( \sum_{k=0}^{2^q - 1} Q_{jk} \psi_k \right) |j\rangle
\label{eq:unitary_action}
\end{equation}

Substituting this into the loss function for each record, we obtain the final QUBO expression:
\begin{equation}
\mathrm{QUBO}_d = \frac{1}{r} \sum_{i=0}^{r - 1} \sum_{j=0}^{2^q - 1} \left( \psi_{ij} - \sum_{k=0}^{2^q - 1} Q_{jk} \psi_{ik} \right)^2
\label{eq:final_qubo}
\end{equation}

This final expression defines a cost function that is quadratic in terms of the parameters encoded in $Q$, and can therefore be expressed as a QUBO formulation. The optimization task of training the variational circuit is thus mapped to finding the binary configuration that minimizes this objective function using a quantum annealer. 

\subsubsection{Quantum Operator representation}\label{sec.quantum_operator} 

A quantum operator is a mathematical entity that represents a physical observable or transformation in a quantum system. It is typically expressed as a matrix acting on a quantum state vector in a Hilbert space, allowing the computation of measurement outcomes and system evolution. These operators follow specific algebraic properties, such as Hermitian symmetry for observables, ensuring real eigenvalues that correspond to measurable quantities.

Here we describe the procedure for constructing a symbolic representation of the Ansatz, to be incorporated into the QUBO formulation as in Equation \ref{eq:final_qubo}. This procedure is provided for the \texttt{TwoLocal} Ansatz as an example, but can be generalized to other ans\"atze as described below.

\begin{enumerate}
    \item \textbf{Extraction of the Quantum Operator}: The Ansatz circuit is mapped to its corresponding quantum operator, expressed symbolically.
    
    \item \textbf{Variable Substitution}: The parameterized rotation angles \( \theta_0, \theta_1, \dots \) are relabeled using symbolic alphabetic variables \( a, b, c, \dots \) to facilitate algebraic manipulation.
    
    \item \textbf{Hyperbolic Reformulation}: The \texttt{TwoLocal} Ansatz generates an operator consisting of exponential functions of the form \( e^{\theta_i} \) and \( e^{-\theta_i} \), which appear in pairs due to the structure of the unitary evolution. These terms are rewritten in terms of hyperbolic functions:
    \begin{equation}
        e^{\theta_i} + e^{-\theta_i} = 2\cosh(\theta_i),
    \end{equation}
    along with analogous transformations for \( -\cosh(\theta_i) \), \( \sinh(\theta_i) \), and \( -\sinh(\theta_i) \).
    
    \item \textbf{Product Expansion of Hyperbolic Functions}: The reformulated operator now consists of products of hyperbolic functions, which can be further decomposed using the expansion:

    \begin{equation}
        \prod_{i=1}^n \cosh(x_i) = \frac{1}{2^n} \sum_{\mathbf{s} \in \{-1,1\}^n} \exp\left( \sum_{i=1}^n s_i x_i \right)
        \label{eq:cosh_product_expansion}
    \end{equation}
    
    This representation transforms the operator into a summation over exponential terms, making it suitable for encoding into a QUBO formulation. Each vector $\mathbf{s} = (s_1, s_2, \dots, s_n)$ is an $n$-dimensional vector whose elements are either $+1$ or $-1$. The total number of such vectors is $2^n$, hence the normalization factor $1/2^n$. However, for full compliance with QUBO encodings, in which variables are constrained to binary values, we substitute \( s_i = 2b_i - 1 \) to derive the following expression: 

    \begin{equation}
        \prod_{i=1}^n \cosh(x_i) = \frac{1}{2^n} \sum_{\mathbf{b} \in \{0,1\}^n} \exp\left( \sum_{i=1}^n (2b_i - 1) x_i \right)
        \label{eq:cosh_expansion_binarized}
    \end{equation}
    
    where $\mathbf{b}$ is a binary vector of length $n$, with $b_i \in \{0,1\}$. 

    A basic example is provided in Appendix \ref{ap:operator_example} for better illustration.

\end{enumerate}

\paragraph{Generalization of the quantum operator expression}

According to the Spectral Theorem, a Hermitian matrix is guaranteed to have a complete set of linearly independent eigenvectors, which means it is diagonalizable. The matrix can be written as:

\begin{equation}
A = U \Lambda U^\dagger
\end{equation}
 
where \( U \) is a unitary matrix (columns are orthonormal eigenvectors of \( A \)) and \( \Lambda \) is a diagonal matrix containing the eigenvalues of \( A \).

%Given this property, if the ansatz operator can be diagonalized, the QUBO formulation can be constructed more efficiently as only one element per row in the matrix needs to be considered in the dot product against each input State Vector (all other row entries will be zero). Each input state vector can be transformed via inner product with the diagonal operator, which simply scales the amplitudes by the eigenvalues and only the measured amplitudes are used to define the final QUBO terms.

Given this property, if the ansatz operator is Hermitian, or can be symmetrized or approximated as such, it admits a diagonal representation that simplifies the construction of the QUBO formulation. In this representation, only one matrix element per row contributes to the dot product with each input state vector, since all off-diagonal elements vanish. As a result, the transformation reduces to scaling each amplitude by the corresponding eigenvalue, and only the resulting measured amplitudes are used to define the final QUBO terms.

The main challenge lies in deriving the diagonal form of a general parameterized unitary operator, which becomes intractable for large systems due to the exponential size of the Hilbert space. Although some operators (like Pauli or diagonal unitaries) allow for simpler diagonalization, general circuits may require advanced methods such as SVD \cite{golub2013matrix} or Lie algebra and spectral graph theory-based techniques \cite{nielsen2010quantum}.

Nevertheless, once a suitable diagonal representation is found, the QUBO construction becomes straightforward: eigenvalues define the cost terms, and the inner products with eigenvectors guide the optimization landscape.

\subsubsection{Updating QUBO formulation and applying global constraints}\label{sec:constraints}

In our QUBO formulation, we discretize the range of each variational angle into a finite number of possible values. To ensure that exactly one value is selected for each angle during optimization, we impose a \textbf{uniqueness constraint} on the corresponding binary variables.

Let $d$ be the number of discretization levels (i.e., the number of segments into which the search range is divided), and let $a$ be the $m$ variational angles in the Ansatz circuit ($a_i$ for $i \in \{0, .., m-1\}$). For each angle $a_i$, we define binary variables $C_{a_{i1}}, C_{a_{i2}}, \dots, C_{a_{id}}$, where each variable indicates whether the angle $a_i$ takes the corresponding discrete value. 

We enforce the following constraint:
\begin{equation}
\sum_{j=1}^{d} C_{a_{ij}} = 1 \qquad \text{for all } a_i , i\in \{0, .., m-1\}
\label{eq:uniqueness_constraint}
\end{equation}

This condition ensures that only one discretized value is selected per angle, which is necessary for the solution to be valid within the QUBO framework. During the search process, the model explores the midpoints of these discretized segments. At each iteration, only configurations satisfying the constraint in Equation~\ref{eq:uniqueness_constraint} (i.e., with exactly one bit activated per angle) are accepted as feasible.

The general process for creating the QUBO expression is summarized in Algorithm \ref{alg:vqa_adiabatic}.

\begin{algorithm}[H]
\caption{QUBO Construction for VQA Training}
\label{alg:vqa_adiabatic}
\begin{algorithmic}[1]
    \State \textbf{Discretize} search space in \textbf{d} partitions
    \For{each record in dataset}
        \State Formulate a \textbf{Quantum Operator representation} (as described in Sec.~\ref{sec.quantum_operator})
        \State Construct a \textbf{QUBO expression for the MSE loss} (as in Sec.~\ref{secformulation})
    \EndFor
    \State Add \textbf{global QUBO constraints} (as in Sec.~\ref{sec:constraints})
\end{algorithmic}
\end{algorithm}

\subsection{Hierarchical Recursive Optimization}\label{sec:optimization}

In the context of combinatorial optimization using VQAs, achieving high precision parameter values in a single execution poses significant computational challenges. Directly optimizing a quantum operator over a continuous search space requires expanding the operator into a large scale formulation, typically represented as a QUBO model. However, the size of the resulting QUBO formulation grows exponentially with the number of binary variables needed to represent fine grained precision, which leads to severe hardware constraints, particularly with respect to qubit count.

To mitigate these limitations, we propose a hierarchical recursive search strategy that enables efficient exploration and exploitation of the parameter space while maintaining a tractable computational cost. This method falls within the family of hierarchical or multilevel metaheuristics~\cite{valejo2021critical, luke2013essentials}, where the search space is iteratively refined through partitioning and selective focus on promising subregions. The approach balances global exploration with local intensification by combining coarse grained search over a broad space and subsequent fine grained optimization around the best found solutions. Conceptually, it also shares similarities with variable neighborhood search techniques~\cite{mladenovic1997variable}.

\begin{figure}
    \centering
    \includegraphics[width=0.6\linewidth]{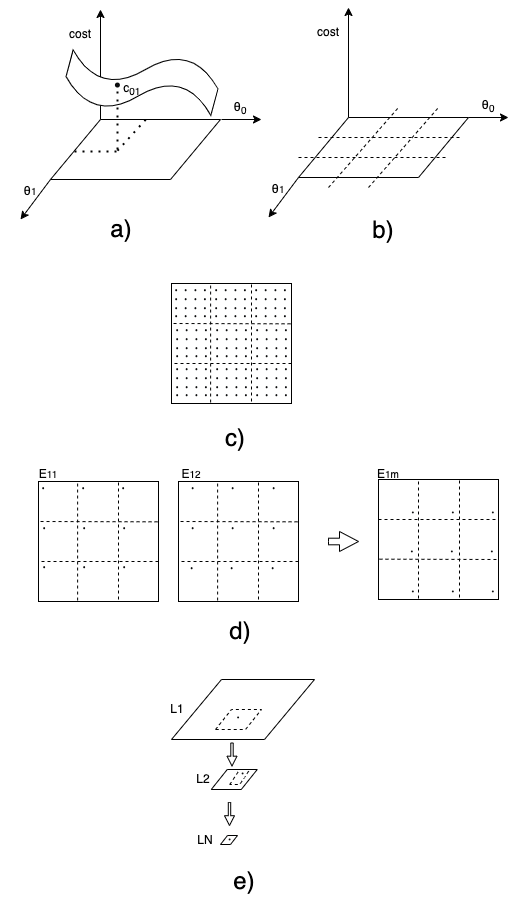}
    \caption{Hierarchical segmentation and discretization. (a) Search space with 2 parameters and cost function. (b) Partitioning of the space into regions. (c) Discretization of each region. (d) Level 1 exploration using evaluation points. (e) Recursive exploitation through levels.}
    \label{fig:iterative-model}
\end{figure}

An overview of the algorithmic workflow is shown in Algorithm~\ref{alg:vqa_recursive} and a basic example for illustration is provided in Appendix ~\ref{ap:training_example}. Initially, the parameter space is discretized into $d$ segments along each dimension, leading to $d^q$ distinct regions in a $q$-dimensional parameter space (Fig.~\ref{fig:iterative-model}.b). Each region is then segmented into $w$ evaluation points per axis, producing $w^q$ candidate solutions per region (Fig.~\ref{fig:iterative-model}.c). In each iteration, a group of evaluation points (consisting of one point from each region at the same relative position) is selected and used to execute the quantum algorithm (Fig.~\ref{fig:iterative-model}.d). These executions yield validation results for each configuration, and the best performing configuration is used as the pivot for refining the search in the subsequent level (Fig.~\ref{fig:iterative-model}.e).

\begin{center}
\begin{minipage}{0.9\textwidth}
\begin{algorithm}[H]
\caption{Hierarchical Recursive Optimization}
\label{alg:vqa_recursive}
\begin{algorithmic}[1]
    \State \textbf{Define} accuracy threshold $\tau$
    \State \textbf{Define} maximum search levels $L$
    \State \textbf{Load} training dataset

    \While{level $< L$ \textbf{and} accuracy drop\textsuperscript{*} $< \tau$}
        \State \textbf{Discretize} parameter space into $d$ segments for each angle $a$
        \State \textbf{Define} $w$ evaluation points per segment
        \For{each combined parameter set (all $w$'s for all $a$'s)}
            \State \textbf{Run} Adiabatic Algorithm with QUBO formulation
            \State \textbf{Get} lowest energy sample \State \textbf{Execute VQA} with corresponding parameters.
            \If{execution accuracy $>$ best accuracy}
                \State \textbf{Update} best accuracy
            \EndIf
            \State Update Parameters-Accuracy map
        \EndFor
        \State \textbf{Return} Parameters for highest accuracy in the map
        \State \textbf{Redefine} region boundaries and discretization level around best\\ Parameters and move to next level $l$.
    \EndWhile
\end{algorithmic}
\end{algorithm}
\vspace{-1.5em}
{\small\textsuperscript{*}\textbf{accuracy drop} is computed as the difference between current and best accuracy.}
\bigskip
\end{minipage}
\end{center}

To enhance statistical robustness and fairness, a randomized distribution of evaluation points is generated prior to the exploration phase, ensuring equal sampling probability across the space. 

Each recursion level adaptively redefines the search space and optionally rescales the number of discretization points. This allows the algorithm to focus computational effort on promising subspaces without revisiting low-potential regions. Moreover, by enabling different configurations of segment counts $d$, discretization granularity $w$, and recursion depth $L$, the model remains highly adaptable to hardware limitations and specific problem requirements.

As a simple example for illustration, consider a VQA with 3 qubits ($q=3$) and 2 parameterized Pauli-Y rotation gates ($a=2$), leading to a 2-parameter system with angles $\theta_0$ and $\theta_1$ defined over $[0, 2\pi)$. Each angle’s domain is partitioned into $d=3$ intervals: $[0, 2\pi/3)$, $[2\pi/3, 4\pi/3)$, $[4\pi/3, 2\pi)$, yielding 9 ($d^a=3^2$) regions (Fig.~\ref{fig:iterative-model}.b). Each region is discretized into $w=4$ points per angle, resulting in $16$ ($w^a=4^2$) points per region evaluated in parallel, and $144$ validation points overall ($(d*w)^2=(3*4)^2$) (Fig.~\ref{fig:iterative-model}.c). %The Adiabatic process is executed once per group of evaluation points, covering all relative positions across regions in 16 executions ($w^2=4^2$). The configuration with the lowest Hamiltonian energy determines the focus of the next recursive iteration.
The adiabatic process is executed once per set of evaluation points with identical relative positions across all regions, requiring \( w^a = 16 \) executions per recursion level (Fig.~\ref{fig:iterative-model}.d).

This approach enables VQAs to efficiently explore high dimensional combinatorial spaces while controlling the growth of the binary variable count in QUBO formulations. By recursively narrowing the parameter space and refining discretizations, the method converges toward optimal solutions without requiring an exhaustive global search. The recursive hierarchical design promotes computational efficiency and model accuracy, balancing scalability with solution precision.

\section{Experiments and Results}\label{experiment}

To evaluate the effectiveness of the proposed methodology, we conducted a comparative analysis across three distinct datasets, employing a range of state-of-the-art methodologies, namely classical optimization techniques alongside evolutionary algorithms to optimize the parameters of VQAs.

In order to compare the performance of the different training methods, the same datasets, ansatz, and system configuration were used. The code is available at: \href{https://github.com/eacostam/gradientfreeVQA.git}{https://github.com/eacostam/gradientfreeVQA.git}.

\subsection{Datasets}

We evaluated the proposed methods on three datasets: \textit{Iris}, \textit{Heart Disease} and \textit{Diabetes} diagnosis. All datasets are publicly available at: \url{https://archive.ics.uci.edu/datasets}

\textit{Iris} dataset is a small, well structured classic dataset with 150 flower samples across three species, each described by four numerical features. Originally collected for discriminant analysis, it’s widely used to validate classification models. In order to test the binary classification models, we work with the registers corresponding to the first two classes, setosa and versicolor. 

\textit{Heart Disease} dataset comprises 76 attributes, though published studies have predominantly utilized a subset of 14 features. Notably, the Cleveland subset is the primary dataset analyzed in machine learning research. The target variable, referred to as the ``goal" field, represents the presence of heart disease and is encoded as an integer ranging from 0 to 4, where 0 indicates the absence of the condition, and values 1, 2, 3, and 4 indicate varying degrees of presence. Research involving the Cleveland dataset has primarily focused on the binary classification task of distinguishing between the presence (1, 2, 3, 4) and absence (0) of heart disease \cite{Detran1989Coronary}. 

Finally, Pima Indian \textit{Diabetes} Dataset, originally from the National Institute of Diabetes and Digestive and Kidney Diseases, contains information of 768 women from a population near Phoenix, Arizona, USA. The outcome tested was Diabetes, 258 tested positive and 500 tested negative. Therefore, there is one target (dependent) variable and 9 attributes \cite{Smith1988ADAP}. 

\subsection{Algorithms and parameter settings}

\begin{figure}[h]
    \centering
    \begin{tikzpicture}
    \begin{axis}[
        boxplot/draw direction=y,
        ylabel={Accuracy},
        xtick={1,2,3},
        xticklabels={ADAM, COBYLA, SPSA},
        width=10cm,
        height=6cm, 
        legend style={at={(0.5,-0.15)}, anchor=north, legend columns=-1}
    ]
    
    \addplot+[
        boxplot,
        fill=blue!20,
        draw=blue
    ] table[y=Value, col sep=comma] {adam.csv};
    
    \addplot+[
        boxplot,
        fill=orange!20,
        draw=orange
    ] table[y=Value, col sep=comma] {cobyla.csv};
    
    \addplot+[
        boxplot,
        fill=green!20,
        draw=green
    ] table[y=Value, col sep=comma] {spsa.csv};
    
    \end{axis}
    \end{tikzpicture}
    \caption{Performance of classical training for VQAs with different optimizers for \textit{Diabetes} dataset.}
    \label{fig:classical_optimizers}
\end{figure}
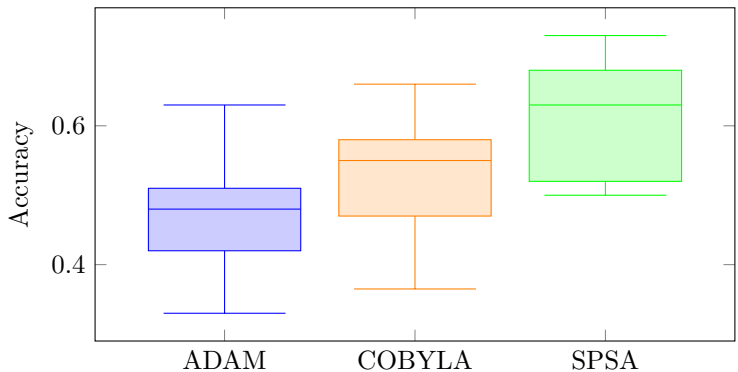

For classical training of the VQAs on the \textit{Diabetes} dataset, we tested different optimization methods: ADAM \cite{kingma2015adam}, SPSA \cite{spall1992spsa} and COBYLA \cite{powell1994cobyla}, having the best performance with SPSA as shown in Figure~\ref{fig:classical_optimizers}. A number of 100 iterations was set for these classical optimizers.  Automatic calibration was used to find the best learning rate and perturbation values according to \cite{Kandala2017VQE}.

In the case of Evolutionary Training, the search algorithm selected was the Memetic Algorithm \cite{moscato1989memetic} with a total of 10 generations, Tournament Selection and Gaussian Mutation with a sigma factor of 0.2 and mutation probability of 15\%.

For the adiabatic training, different configurations of the model were used in terms of levels, search space discretization and number of validation points.

\subsection{Data Preparation}

For the \textit{Diabetes} and \textit{Heart Disease} datasets, Principal Component Analysis (PCA) is used to reduce the number of features to match the number of qubits in the circuit.

In order to maintain proper balance in the training and validation data sets, oversampling and under sampling techniques were used. We implemented SMOTE strategies to increase the number of samples in the class with fewer records based on the existing ones, and Random Sampling in the cases under sampling is selected.

Shuffling was used in the experiments prior each training execution in order to avoid learning bias as much as possible.

%Dataset splitting was performed so that a 80/20/20 distribution has been applied on the dataset for convenient cross validation.  That is, 
The original datasets are split in 80\% for training+testing and 20\% for validation, then the first subsets are split again in 80\% records for training and 20\% for testing.

\subsection{Experimental evaluation of the proposed model}

The proposed model allows for different configuration of parameters, adjustable to the specific problem, leading to different accuracies and execution times as shown in Figure \ref{fig:pareto}. Complete results tables are available in Appendix \ref{ap:parameter_exploration_data}. For the \textit{Diabetes} dataset, the best performance/cost ratio is reached with parameter settings $H$ (2 Partitions, 3 validation points) and $D$ (3 Partitions, 3 Validation points).  In the case of \textit{Heart Disease}, the best performance/cost ratio is reached with parameter setting $B$ (3 Partitions, 3 Validation Points).  For the \textit{Iris} dataset, the best training configuration is 2 Partitions, 1 Validation Points (configuration $I$), and 2 Partitions, 2 Validation Points (configuration $J$).

\begin{figure}[H]
\centering

% First plot (Diabetes) 
\begin{minipage}{0.7\textwidth}
\centering
\begin{tikzpicture}
\begin{axis}[
    width=\linewidth,
    height=6cm,
    xlabel={Training Accuracy},
    ylabel={Training Time},
    title={\textit{Diabetes} Dataset},
    grid=major,
    legend pos=north west,
    ymin=0,
    enlargelimits=0.05,
    scaled y ticks=false,
    nodes near coords,
    point meta=explicit symbolic,
    every node near coord/.append style={
        font=\tiny\bfseries,
        inner sep=0pt,
        text=white,
        align=center,
        anchor=center
    }
]
% All points (gray)
\addplot[
    only marks,
    mark=*,
    mark size=4pt,
    color=gray
] table[
    col sep=comma,
    x=taccuracy,
    y=time,
    meta=label
] {csv/diabetes_results_detailed_avgs.csv};

% Pareto front
\addplot[
    thick,
    color=red,
    mark=*,
    mark size=4pt,
    visualization depends on={value \thisrow{label} \as \mylabel},
    nodes near coords=\mylabel,
    point meta=explicit symbolic
] table [meta=label] {
    taccuracy  time     label
    0.5105     14.122   O
    0.6125     14.253   J
    0.625      7.008    D
    0.6375     14.15    G
    0.65       6.93     A
    0.658      21.51    O
    0.675      28.77    U
    0.6875   104.454    B1
    0.6975   106.901    B
    0.7125   514.866    H
    0.72625  1644.08    D
    0.7333   4107.24    E
};
\end{axis}
\end{tikzpicture}
\end{minipage}
\hfill

% Second plot (Heart Disease)
\begin{minipage}{0.7\textwidth}
\centering
\begin{tikzpicture}
\begin{axis}[
    width=\linewidth,
    height=6cm,
    xlabel={Training Accuracy},
    ylabel={Training Time},
    title={\textit{Heart Disease} Dataset},
    grid=major,
    legend pos=north west,
    ymin=0,
    enlargelimits=0.05,
    scaled y ticks=false,
    nodes near coords,
    point meta=explicit symbolic,
    every node near coord/.append style={
        font=\tiny\bfseries,
        inner sep=0pt,
        text=white,
        align=center,
        anchor=center
    }
]
% All points (gray)
\addplot[
    only marks,
    mark=*,
    mark size=4pt,
    color=gray
] table[
    col sep=comma,
    x=taccuracy,
    y=time,
    meta=label
] {csv/heartdisease_results_detailed_avgs.csv};

% Pareto front
\addplot[
    thick,
    color=red,
    mark=*,
    mark size=4pt,
    visualization depends on={value \thisrow{label} \as \mylabel},
    nodes near coords=\mylabel,
    point meta=explicit symbolic
] table [meta=label] {
    taccuracy  time     label
    0.5298     14.18      O
    0.6463     373.7267   A
    0.6760     630.3367   B
    0.6815   16787.15     N
    0.6815   17333.58     M
};
\end{axis}
\end{tikzpicture}
\end{minipage}

% Third plot (Iris Disease)
\begin{minipage}{0.7\textwidth}
\centering
\begin{tikzpicture}
\begin{axis}[
    width=\linewidth,
    height=6cm,
    xlabel={Training Accuracy},
    ylabel={Training Time},
    title={\textit{Iris} Dataset},
    grid=major,
    legend pos=north west,
    ymin=0,
    enlargelimits=0.05,
    scaled y ticks=false,
    nodes near coords,
    point meta=explicit symbolic,
    every node near coord/.append style={
        font=\tiny\bfseries,
        inner sep=0pt,
        text=white,
        align=center,
        anchor=center
    }
]

% All points (gray)
\addplot[
    only marks,
    mark=*,
    mark size=5pt,
    color=gray
] table[
    col sep=comma,
    x=taccuracy,
    y=time,
    meta=label
] {csv/iris_results_detailed_avgs.csv};

% Pareto front manually computed
\addplot[
    thick,
    color=red,
    mark=*,
    mark size=6pt,
    visualization depends on={value \thisrow{label} \as \mylabel},
    nodes near coords=\mylabel,
    point meta=explicit symbolic,
    every node near coord/.append style={
        font=\footnotesize\bfseries,
        inner sep=0pt,
        text=white,
        align=center,
        anchor=center
    }
] table [meta=label] {
    taccuracy  time     label
    0.4625    2.05      O
    0.7708    93.783    A
    0.890625  115.795   K
    0.906     114.03    J
    0.937     115.67    I
};

\end{axis}
\end{tikzpicture}
\end{minipage}

\caption{Average Training Accuracy vs. Time for runs with different parameter settings for the three datasets under consideration. Pareto Front configurations are colored in red. Solution labeled as \textit{O} represents the baseline configuration (one level, one partition, and one point), which corresponds to a classical brute-force search scenario. Complete results available in Appendix \ref{ap:parameter_exploration_data}. }
\label{fig:pareto}
\end{figure}
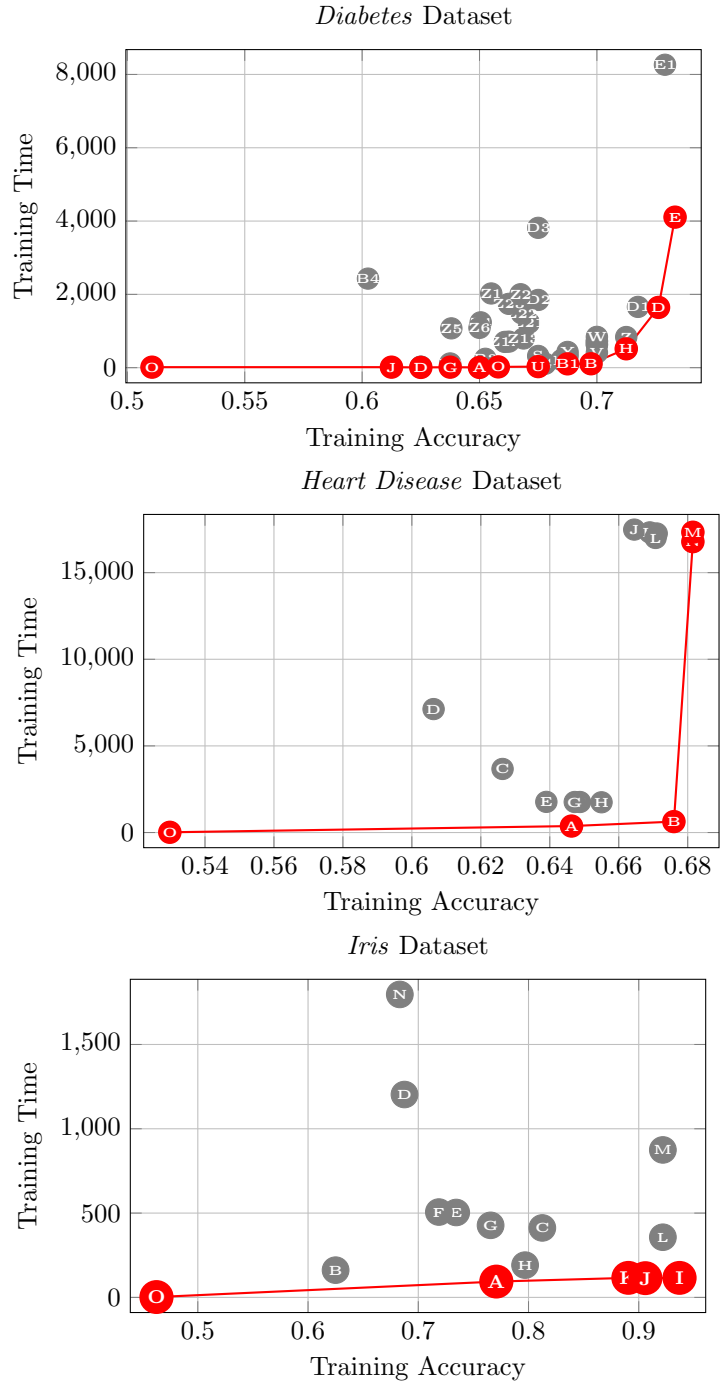

As illustrated in the complete data results table and plots in~Appendixes~\ref{ap:parameter_exploration} and ~\ref{ap:parameter_exploration_data}, the values of the model's parameters significantly impact the obtained results, probably in correspondence with the complexity of the problem. For the \textit{Iris} dataset, a lower number of exploration levels (1) and partitions (2) is sufficient to achieve high accuracy with minimal computational cost, as evidenced by configurations \textit{J}, \textit{K}, and \textit{I}. In the case of the \textit{Diabetes} dataset, configurations \textit{B} and \textit{H}, characterized by a low number of levels (1) and a moderate number of partitions (3), show a significant boost of performance compared to configurations such as \textit{J} and \textit{D}, which use a medium number of levels (2) and a smaller number of partitions (2). Regarding the \textit{Heart Disease} dataset, configurations \textit{A} and \textit{B}, which employ higher exploration levels (3–4) and moderate partitioning and validation point settings (2–3), result in superior accuracy and reduced execution time. In contrast, configuration \textit{M}, with a high number of validation points (5), incurs substantial computational cost without yielding accuracy improvements. Similarly, configuration \textit{O}, which uses minimal levels, partitions, and validation points (all set to 1), leads to a marked accuracy decline exceeding 15 percentage points.  

\subsection{Comparative evaluation of results}

\begin{figure}[H]
    \centering
    \begin{tikzpicture}
        \begin{groupplot}[
            group style={group size=3 by 1, horizontal sep=2cm},
            width=0.32\textwidth,
            height=5cm,
            xlabel={Dataset Size},
            ylabel={Accuracy},
            tick label style={font=\small},
            grid=major,
            legend style={at={(2.5,-0.35)}, anchor=north, legend columns=-1}
        ]

        % First plot - Iris dataset
        \nextgroupplot[title={\textit{Iris}}]
            
            \addplot [color=blue, mark=*,
        mark options={fill=blue}] table [x=size, y=ad_acc, col sep=comma] {allmodels_iris122.csv};
            \addlegendentry{Adiabatic}

            \addplot [color=red, mark=*,
        mark options={fill=red}] table [x=size, y=ev_acc, col sep=comma] {allmodels_iris121.csv};
            \addlegendentry{Evolutionary}

            \addplot [color=brown, mark=*,
        mark options={fill=brown}] table [x=size, y=cl_acc, col sep=comma] {allmodels_iris121.csv};
            \addlegendentry{Classical}

        % Second plot - Heart Disease dataset
        \nextgroupplot[title={\textit{Heart Disease}}]

            \addplot [color=blue, mark=*,
        mark options={fill=blue}] table [x=size, y=ad_acc, col sep=comma] {allmodels_heart133.csv};
            \addplot [color=red, mark=*,
        mark options={fill=red}] table [x=size, y=ev_acc, col sep=comma] {allmodels_heart123.csv};
            \addplot [color=brown, mark=*,
        mark options={fill=brown}] table [x=size, y=cl_acc, col sep=comma] {allmodels_heart123.csv};

        % Third plot - Diabetes dataset
        \nextgroupplot[title={\textit{Diabetes}}]
        
            \addplot [color=blue, mark=*,
        mark options={fill=blue}] table [x=size, y=ad_acc, col sep=comma] {allmodels_diabetes123.csv};
            
            \addplot [color=red, mark=*,
        mark options={fill=red}] table [x=size, y=ev_acc, col sep=comma] {allmodels_diabetes123.csv};
            
            \addplot [color=brown, mark=*,
        mark options={fill=brown}] table [x=size, y=cl_acc, col sep=comma] {allmodels_diabetes123.csv};
        \end{groupplot}
    \end{tikzpicture}
    \caption{Validation accuracy across datasets}
    \label{fig:accuracies}
\end{figure}
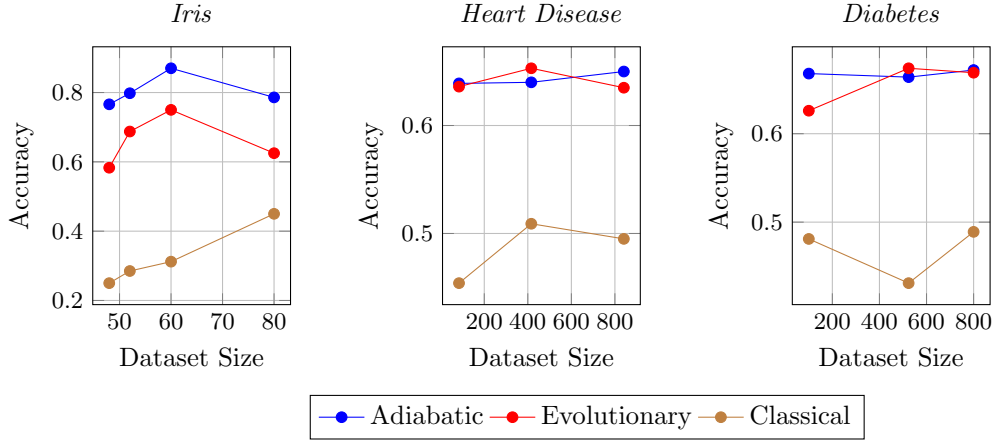

\begin{figure}[H]
    \centering
    \begin{tikzpicture}
        \begin{groupplot}[
            group style={group size=3 by 1, horizontal sep=2cm},
            width=0.32\textwidth,
            height=5cm,
            xlabel={Dataset Size},
            ylabel={Time execution},
            tick label style={font=\small},
            grid=major,
            legend style={at={(2.5,-0.35)}, anchor=north, legend columns=-1}
        ]

        % First plot - Iris dataset
        \nextgroupplot[title={\textit{Iris}}]

        \addplot [color=blue, mark=*,
        mark options={fill=blue}] table [x=size, y=ad_t, col sep=comma] {allmodels_iris122.csv};
            \addlegendentry{Adiabatic}

            \addplot [color=red, mark=*,
        mark options={fill=red}] table [x=size, y=ev_t, col sep=comma] {allmodels_iris121.csv};
            \addlegendentry{Evolutionary}

            \addplot [color=brown, mark=*,
        mark options={fill=brown}] table [x=size, y=cl_t, col sep=comma] {allmodels_iris121.csv};
            \addlegendentry{Classical}

        % Second plot - Heart Disease dataset
        \nextgroupplot[title={\textit{Heart Disease}}]

            \addplot [color=blue, mark=*,
        mark options={fill=blue}] table [x=size, y=ad_t, col sep=comma] {allmodels_heart133.csv};
            \addplot [color=red, mark=*,
        mark options={fill=red}] table [x=size, y=ev_t, col sep=comma] {allmodels_heart123.csv};
            \addplot [color=brown, mark=*,
        mark options={fill=brown}] table [x=size, y=cl_t, col sep=comma] {allmodels_heart123.csv};

        % Third plot - Diabetes dataset
        \nextgroupplot[title={\textit{Diabetes}}]

            \addplot [color=blue, mark=*,
        mark options={fill=blue}] table [x=size, y=ad_t, col sep=comma] {allmodels_diabetes123.csv};
            \addplot [color=red, mark=*,
        mark options={fill=red}] table [x=size, y=ev_t, col sep=comma] {allmodels_diabetes123.csv};
            \addplot [color=brown, mark=*,
        mark options={fill=brown}] table [x=size, y=cl_t, col sep=comma] {allmodels_diabetes123.csv};
        \end{groupplot}
    \end{tikzpicture}
    \caption{Time execution across datasets}
    \label{fig:time_executions}
\end{figure}
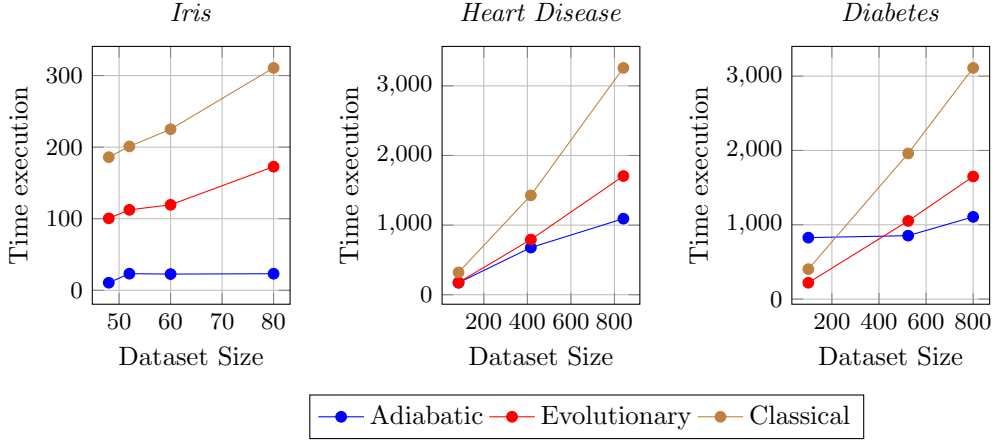

In order to assess the generalization capability and computational cost of the proposed methodology, in this section we perform a comparative evaluation of results with respect to classical and evolutionary optimization strategies for the three datasets under consideration. In this comparison, we focus on the validation accuracies and running times for the different approaches with datasets of increasing size for each problem, as illustrated in Figures~\ref{fig:accuracies} and~\ref{fig:time_executions}. 
Particularly, Figure~\ref{fig:accuracies} shows that the proposed Adiabatic Quantum Machine Learning approach achieves superior validation performance compared to Classical and Evolutionary methods, particularly on small dataset sizes. This observation suggests that the proposed model could achieve superior generalization performance in data-constrained settings.

A key advantage of the Adiabatic approach lies in its computational efficiency. As shown in Figure~\ref{fig:time_executions}, the entire training process consistently requires less execution time across all datasets compared to the other two models, as the empirical computational complexity scales sublinearly with the number of training records. 

\subsection{Scalability Analysis by Model Parameters}\label{scalability}

The model is governed by a set of parameters that control its granularity, number of executions, and balance between exploration and exploitation of the search space. These parameters impact the model’s scalability as summarized below:

\begin{itemize}
\item The number of terms in the MSE cost function scales linearly with the number of training records ($r \cdot f$).
\item The number of binary variables in the QUBO formulation grows linearly with the number of circuit angles and partitions ($a \cdot d$).
\item The degree of parallelism increases exponentially with the number of partitions and angles ($d^a$).
\item The number of training executions scales exponentially with the number of validation points and angles ($w^a$).
\end{itemize}

Where $a$ denotes the number of parameterized angles in the VQA circuit, $f$ the number of QUBO terms per record, $r$ the number of training records, $d$ the number of search space partitions, and $w$ the number of validation points per angle. 

Different experiments have been carried out to measure the empirical efficiency of the proposed model as the number of training records, partitions, validation points and levels increase for the different datasets under consideration. These results are available in Appendixes~\ref{ap:parameter_exploration} and~\ref{ap:parameter_exploration_data} and are in concordance with the previous statements.  

\paragraph{Search space partitioning}

With respect to search space partitioning and the balance between exploration and exploitation, Figure \ref{fig:time_executions} shows that computational cost tends to scale superlinearly, potentially following a polynomial trend, depending on the dataset characteristics and the interplay between partitioning granularity and search efficiency. 

The model's complexity is likely governed by the increasing number of function evaluations required as partitions grow (see for reference Figures \ref{fig:diabetes_performance} and \ref{fig:diabetes_training_times} for the observed on the \textit{Diabetes} dataset experiments), which is a known challenge in high dimensional search spaces, particularly in hybrid classical-quantum or combinatorial optimization frameworks.

Therefore, it’s important to strike a careful balance between computational cost and accuracy. While increasing partitioning allows for a more detailed search of the solution space, it might also reduce accuracy. To address this, adaptive partitioning strategies or hybrid approaches may be needed to manage the added computational burden. 

\begin{figure}[H]
    \centering
    \begin{tikzpicture}
        \begin{groupplot}[
            group style={
                group size=2 by 1,
                horizontal sep=2cm,
            },
            width=6cm,
            height=5cm,
            xlabel={Search level (l)},
            xtick=data,
            xmin=0.8, xmax=4.2, 
            tick label style={align=center},
            grid=major,
        ]

        % Plot 1: Accuracy
        \nextgroupplot[
          ylabel={Accuracy},
          legend style={font=\small, at={(1.2,-0.3)}, anchor=north, legend columns=3}
        ]
        \addplot table [x=levels, y=acc, col sep=comma] {adiabatic_scaling_levels_iris.csv};
        \addlegendentry{Iris}
        
        \addplot table [x=levels, y=acc, col sep=comma] {adiabatic_scaling_levels_heart_disease.csv};
        \addlegendentry{Heart Disease}
        
        \addplot table [x=levels, y=acc, col sep=comma] {adiabatic_scaling_levels_diabetes.csv};
        \addlegendentry{Diabetes}

        % Plot 2: Time
        \nextgroupplot[
          ylabel={Time per execution},
        ]

        \addplot table [x=levels, y expr=\thisrow{t} / (\thisrow{ds_length} * \thisrow{iterations}), col sep=comma] {adiabatic_scaling_levels_iris.csv};
        %\addlegendentry{Iris}
        
        \addplot table [x=levels, y expr=\thisrow{t} / \thisrow{ds_length} * \thisrow{iterations}, col sep=comma] {adiabatic_scaling_levels_heart_disease.csv};
        %\addlegendentry{Heart Disease}
        
        \addplot table [x=levels, y expr=\thisrow{t} / \thisrow{ds_length} * \thisrow{iterations}, col sep=comma] {adiabatic_scaling_levels_diabetes.csv};
        %\addlegendentry{Diabetes}
        
        \end{groupplot}
    \end{tikzpicture}
    \caption{Performance scaling across datasets for different search space levels ($l$)}
    \label{fig:adiab_scaling_levels_accuracy}
\end{figure}
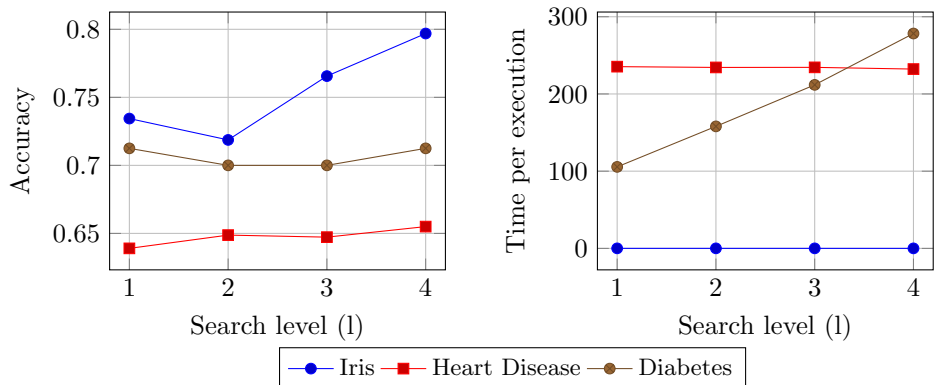

\paragraph{Depth of the search space}
In terms of depth of search space, the computational cost of the model increases in a polynomial or higher fashion with search depth, while accuracy gains exhibit logarithmic-like diminishing returns (see Figure \ref{fig:adiab_scaling_levels_accuracy}). This behavior aligns with theoretical results in combinatorial optimization, where deeper searches require exponentially more resources but do not necessarily provide proportionate accuracy improvements. Future refinements could focus on adaptive heuristics to balance efficiency and accuracy. 

\subsection{Error characterization} \label{sec.error_characterization}

Previous studies in classical ML show that controlled noise injection can improve generalization. More recently, quantum noise has also been found to smooth the loss landscape in QML tasks~\cite{bagaev2025, kashif2024hqnet}. 

\begin{figure}[H]
    \centering
    \begin{tikzpicture}
      \begin{axis}[
        boxplot/draw direction=y,
        boxplot={ box extend=0.18 },
        boxplot/every median/.style={solid},
        boxplot/every box/.style={solid},
        boxplot/every whisker/.style={solid},
        boxplot/every average/.style={solid},
        ylabel={Accuracy},
        xlabel={Noise level},
        xtick={1,1.8,2.4},
        xticklabels={$0.0$, $0.1$, $0.2$},
        xmin=0.5,
        xmax=3.0,
        width=13cm,
        height=7cm,
        legend style={font=\small, at={(0.5,-0.3)}, anchor=north, legend columns=3},
        legend image post style={sharp plot, mark=*, mark options={fill=none}, line width=1pt}
      ]
    
      % noise = 0.0
      \addplot+[
        boxplot,
        fill=orange!20,
        draw=orange,
        boxplot/draw position=0.8,
      ] table[y=Value, col sep=comma, header=true] {noise_accu_iris_00.csv};
      \addlegendentry{Iris}
    
      \addplot+[
        boxplot,
        fill=red!20,
        draw=red,
        boxplot/draw position=1.00,
      ] table[y=Value, col sep=comma, header=true] {noise_accu_diabetes_00.csv};
      \addlegendentry{Diabetes}
    
      \addplot+[
        boxplot,
        fill=blue!20,
        draw=blue,
        boxplot/draw position=1.2,
      ] table[y=Value, col sep=comma, header=true] {noise_accu_heartd_00.csv};
      \addlegendentry{Heart Disease}
    
      % noise = 0.1
      \addplot+[
        boxplot,
        fill=orange!20,
        draw=orange,
        boxplot/draw position=1.6,
      ] table[y=Value, col sep=comma, header=true] {noise_accu_iris_01.csv};
    
      \addplot+[
        boxplot,
        fill=red!20,
        draw=red,
        boxplot/draw position=1.8,
      ] table[y=Value, col sep=comma, header=true] {noise_accu_diabetes_01.csv};
    
      \addplot+[
        boxplot,
        fill=blue!20,
        draw=blue,
        boxplot/draw position=2.0,
      ] table[y=Value, col sep=comma, header=true] {noise_accu_heartd_01.csv};
    
      % noise = 0.2
      \addplot+[
        boxplot,
        fill=orange!20,
        draw=orange,
        boxplot/draw position=2.4,
      ] table[y=Value, col sep=comma, header=true] {noise_accu_iris_02.csv};
    
      \addplot+[
        boxplot,
        fill=red!20,
        draw=red,
        boxplot/draw position=2.6,
      ] table[y=Value, col sep=comma, header=true] {noise_accu_diabetes_02.csv};
    
      \addplot+[
        boxplot,
        fill=blue!20,
        draw=blue,
        boxplot/draw position=2.8,
      ] table[y=Value, col sep=comma, header=true] {noise_accu_heartd_02.csv};
      \end{axis}
    \end{tikzpicture}
    \caption{Noise impact on Accuracy}
    \label{fig:noise_accuracy}
\end{figure}
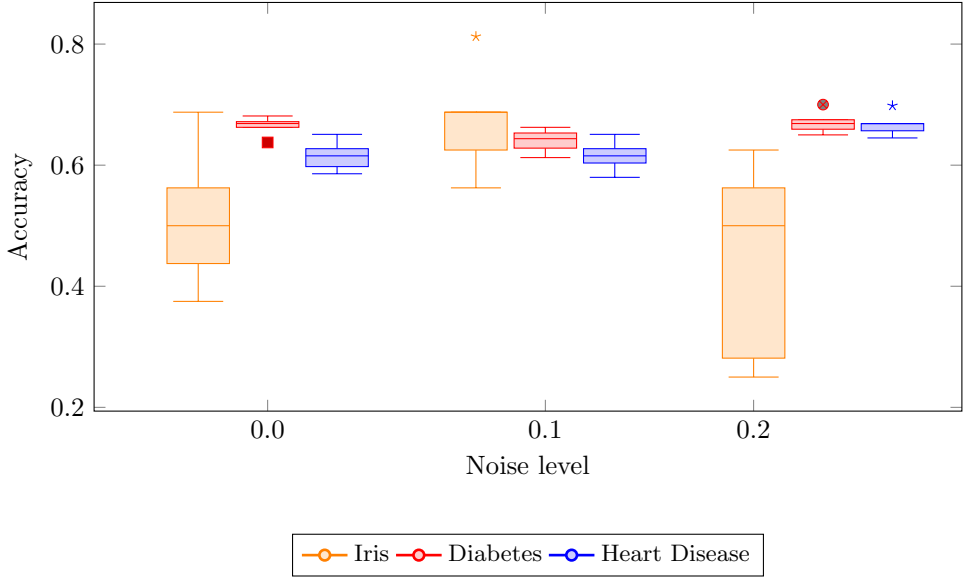

To measure the impact of quantum noise in the performance of the results obtained with the proposed methodology, we perform quantum annealing simulations with varying noise levels. The observed effects are shown in Figure~\ref{fig:noise_accuracy} and 
align with previous findings~\cite{kashif2024hqnet}. For small noise levels (10\%), the performance was generally improved, particularly for \textit{Heart Disease} and the \textit{Iris} datasets. However, higher noise levels (20\%) might be beneficial (\textit{Diabetes}, \textit{Heart Disease}) but also harmful (\textit{Iris}), likely depending on the complexity of the task and sensitivity of the model. As illustrated in the figure, noise levels of 10\% and 20\% occasionally enable the model to attain accuracy values exceeding the mean, which are represented as outliers in the graph.

\section{Conclusions}\label{conclusions}

In this work, we introduced a novel methodology for mapping the optimization problem underlying variational quantum algorithms into a QUBO formulation, which we subsequently address using an adiabatic quantum annealer. %This approach avoids the search for the optimal parameters via gradient descent, and instead uses a stochastic search powered by the entanglement and tunnel effect. 
This approach avoids gradient-based optimization and instead performs parameter search using quantum annealing, which relies on quantum tunneling.

The proposed model was compared to classical and evolutionary VQA optimization strategies on three classical datasets. While our method does not necessarily outperform pure classical machine learning approaches, it demonstrates advantages in either achieving comparable performance with significantly reduced training effort or outperforming classical methods in terms of accuracy. Building on this, our hierarchical method improves training efficiency by refining solutions without increasing memory demands, guided by accuracy thresholds rather than hardware limits. It provides a rich set of configurable parameters, enabling adaptation to specific problem characteristics and available hardware resources. In particular, the model supports customization of the exploration–exploitation balance, degree of adiabatic parallelization, search space discretization granularity, as well as methodologies for adapting to different ansätze. 

However, random initialization in VQAs can introduce bias by confining the search to specific subregions of the parameter space, potentially limiting global exploration. This can hinder convergence to optimal solutions, making structured or informed initialization strategies more effective in achieving better performance. Further work is also required to assess the performance of the proposed methodology on more complex classical problems and inherently quantum problems.

\bmhead{Acknowledgements}

We want to acknowledge funding from PID2021-128970OA-I00 10.13039/501100011033 and PID2021-123041OB-I00 by MCIN/AEI/10.13039/501100011033/, Spanish Government, "FEDER Una manera de hacer Europa". Also the FEDER/Junta de Andalucía program A.FQM.752.UGR20. Finally, we are also grateful for the technical support and computing facilities provided by PROTEUS, the supercomputing center of the Institute Carlos I for Theoretical and Computational Physics in Granada, Spain.

%\section*{Declarations}

%\bibliographystyle{unsrt}
\bibliography{qaa4vqa}% common bib file
%\clearpage
\begin{appendices}

%\clearpage
\newpage
\section{Quantum Operator representation example}\label{ap:operator_example}

\begin{figure}[H]
    \centering
    \includegraphics[width=0.3\linewidth]{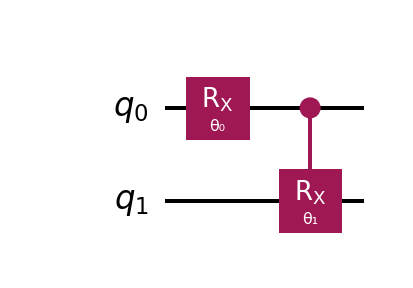}
    \caption{Example circuit}
    \label{fig:example_circuit}
\end{figure}

A simple circuit of two qubits, with two gates, one RX gate acting on the first qubit, and a two qubit RX gate acting on the second qubit with control on the first qubit as seen in Figure \ref{fig:example_circuit}.

The corresponding quantum operator:
\begin{equation}
\begin{bmatrix}
\cos\left(\frac{\theta_0}{2}\right) & -i \sin\left(\frac{\theta_0}{2}\right) & 0 & 0 \\
-i \sin\left(\frac{\theta_0}{2}\right)\cos\left(\frac{\theta_1}{2}\right) & \cos\left(\frac{\theta_0}{2}\right)\cos\left(\frac{\theta_1}{2}\right) & -\sin\left(\frac{\theta_0}{2}\right)\sin\left(\frac{\theta_1}{2}\right) & -i \sin\left(\frac{\theta_1}{2}\right)\cos\left(\frac{\theta_0}{2}\right) \\
0 & 0 & \cos\left(\frac{\theta_0}{2}\right) & -i \sin\left(\frac{\theta_0}{2}\right) \\
-\sin\left(\frac{\theta_0}{2}\right)\sin\left(\frac{\theta_1}{2}\right) & -i \sin\left(\frac{\theta_1}{2}\right)\cos\left(\frac{\theta_0}{2}\right) & -i \sin\left(\frac{\theta_0}{2}\right)\cos\left(\frac{\theta_1}{2}\right) & \cos\left(\frac{\theta_0}{2}\right)\cos\left(\frac{\theta_1}{2}\right)
\end{bmatrix}
\end{equation}

Which can be expressed in the following exponential form (shown only first two columns for easier reading):

\begin{equation}
\begin{bmatrix}
\frac{e^{i\theta_0/2} + e^{-i\theta_0/2}}{2} & -i \cdot \frac{e^{i\theta_0/2} - e^{-i\theta_0/2}}{2i} & \cdots \\
-\frac{e^{i\theta_0/2} - e^{-i\theta_0/2}}{2} \cdot \frac{e^{i\theta_1/2} + e^{-i\theta_1/2}}{2} & \frac{e^{i\theta_0/2} + e^{-i\theta_0/2}}{2} \cdot \frac{e^{i\theta_1/2} + e^{-i\theta_1/2}}{2} & \cdots \\
0 & 0 &\cdots \\
-\frac{e^{i\theta_0/2} - e^{-i\theta_0/2}}{2i} \cdot \frac{e^{i\theta_1/2} - e^{-i\theta_1/2}}{2i} & -i \cdot \frac{e^{i\theta_1/2} - e^{-i\theta_1/2}}{2i} \cdot \frac{e^{i\theta_0/2} + e^{-i\theta_0/2}}{2} &\cdots\\
\end{bmatrix}
\label{mat:op_exp}
\end{equation}

For simplicity, take element [1,1]:
\[
\frac{e^{i\theta_0/2} + e^{-i\theta_0/2}}{2} \cdot \frac{e^{i\theta_1/2} + e^{-i\theta_1/2}}{2}
\]

To facilitate the QUBO encoding, we consider a heuristic transformation of the unitary matrix elements by replacing \( i\theta \rightarrow \theta \), leading to real-valued expressions involving hyperbolic functions.

For instance:

\[
\cos\left( \frac{\theta}{2} \right) = \frac{e^{i\theta/2} + e^{-i\theta/2}}{2}
\quad \longrightarrow \quad
\cosh\left( \frac{\theta}{2} \right) = \frac{e^{\theta/2} + e^{-\theta/2}}{2}
\]

%While this transformation does not preserve unitarity, it enables a direct mapping into QUBO-friendly exponential terms. 
To make the QUBO encoding tractable, we apply a heuristic simplification that replaces the imaginary phase \( i\theta \) with a real-valued angle \( \theta \). Although this substitution breaks unitarity, it transforms complex exponentials into hyperbolic functions, allowing for direct encoding into real-valued QUBO expressions.

For $cosh$ products we can use equation \ref{eq:cosh_expansion_binarized} to derive the following identity:

    \[
    \cosh\left( \frac{\theta_0}{2} \right)\cosh\left( \frac{\theta_1}{2} \right)
    = \frac{1}{4} \sum_{b_0, b_1 \in \{0,1\}} \exp\left( \frac{(2b_0 - 1)\theta_0 + (2b_1 - 1)\theta_1}{2} \right)
    \]

Applied to equation \ref{eq:cosh_product_expansion} becomes: 
\[
\frac{1}{4} \left( e^{\frac{\theta_0 + \theta_1}{2}} + e^{\frac{\theta_0 - \theta_1}{2}} + e^{\frac{-\theta_0 + \theta_1}{2}} + e^{-\frac{\theta_0 + \theta_1}{2}} \right)
\]

The resulting expression consists of sums of exponential terms that can be incorporated into a QUBO formulation, as opposed to the products of exponentials originally derived from the matrix representation \ref{mat:op_exp}.

\FloatBarrier
\clearpage
\newpage
\section{Training process Example}\label{ap:training_example}

\begin{figure}[h]
    \centering
    \includegraphics[width=0.6\linewidth]{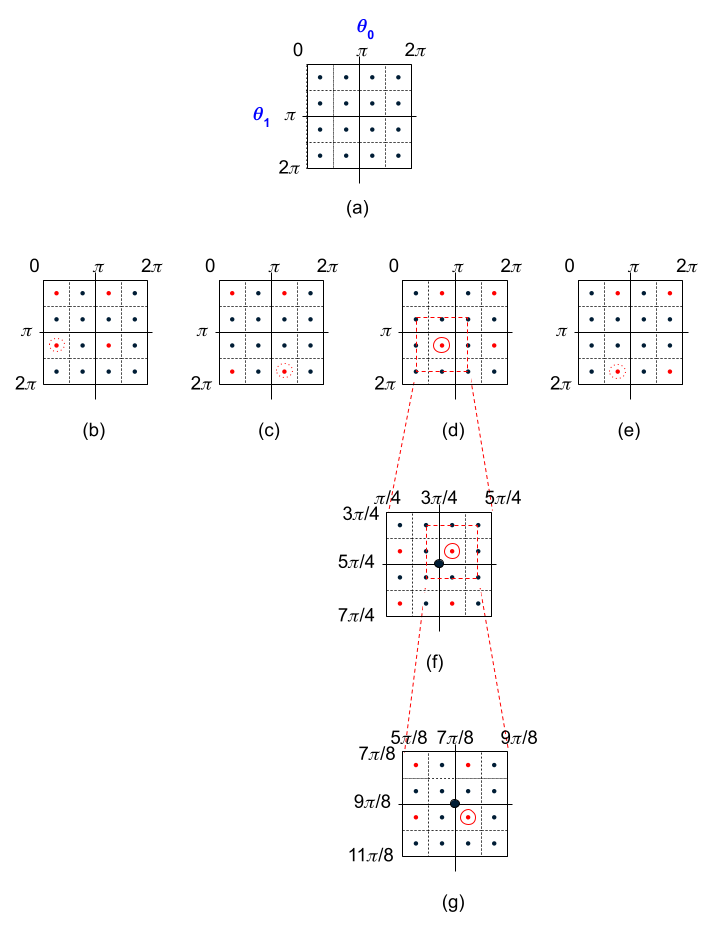}
    \caption{Basic training example. (a) Search space for a 2 dimensional problem discretized in 2 partitions with 2 validation points per angle and partition.  (b) Training Level 1, execution 1. (c) Training Level 1, execution 2. (d) Training Level 1, execution 3. (d) Training Level 1, execution 4. (f) Training Level 2, execution 16. (g) Training Level 3, execution 9.}
    \label{fig:example_training}
\end{figure}

Figure~\ref{fig:example_training} illustrates the recursive optimization process described in Section~\ref{sec:optimization}. 

Based on circuit defined in Appendix \ref{ap:operator_example}, composed by 2 qubits and 2 rotational gates (ie. 2 parameterizable angles: $\theta_0, \theta_1$), let us define the model metaheuristics parameters as follows:

\vspace{1em}

\underline{\textbf{Model Parameters}.  See Figure \ref{fig:example_training} (a)}
\begin{itemize}
    \item Training Levels ($s$): 3
    \item Partitions ($d$): 2
    \item Validation Points per partition and angle ($w$): 2
    \item Angle range rescaling per level: \(50\%\)
\end{itemize}

Leading to:
\begin{itemize}
%    \item QUBO Binary variables ($d*q$): 4
    \item QUBO Binary variables ($d*a$): 4
    \item Validation points evaluated per QUBO execution ($d^q$): 4
    \item Training executions per level ($w^q$): 4
    \item Maximum total executions ($l*w^q)$: 12
\end{itemize}

\vspace{1em}

\underline{\textbf{Training Level 1}}, (Executions 1 to 4):
\begin{itemize} 
    \item \textbf {Angle range}: \(2\pi\)
    \item \textbf{Centroid}: \(\theta_0=\pi\); \(\theta_1=\pi\).
    \item \textbf{Search space}: \(\theta_0:[0 - 2\pi]\); \(\theta_1:[0 - 2\pi]\).
    \item \textbf{Partitions ($d$)}: \(\theta_0:\{[0 - \pi], [\pi - 2\pi]\}\); \(\theta_1:\{[0 - \pi], [\pi - 2\pi]\}\).
    \item \textbf{Validation points ($w$)}: \(\theta_0:\{\pi/4, 5\pi/4\}, \{3\pi/4, 7\pi/4\}\);\\ \(\theta_1:\{\pi/4, 5\pi/4\}, \{3\pi/4, 7\pi/4\}\)
        
    \item \textbf{Training Execution 1}.  Figure \ref{fig:example_training} (b)
    \begin{itemize}
        \item \textbf{Validate for}: \(\theta_0:\{\pi/4, 5\pi/4\}, \theta_1:\{\pi/4, 5\pi/4\}\).
        \item \textbf{Lowest energy solution}: \(\theta_0 = \pi/4\), \(\theta_1 = 5\pi/4\). \textbf{Accuracy}: 55\%.  
        \item \textbf{Best result}: Execution 1
    \end{itemize}

    \item \textbf{Training Execution 2}.  Figure \ref{fig:example_training} (c)
    \begin{itemize}
        \item \textbf{Validate for}: \(\theta_0:\{\pi/4, 5\pi/4\}, \theta_1:\{3\pi/4, 7\pi/4\}\).
        \item \textbf{Lowest energy solution}: \(\theta_0 = 5\pi/4\), \(\theta_1 = 7\pi/4\). \textbf{Accuracy}: 60\%. 
        \item \textbf{Best result}: Execution 2
    \end{itemize}

    \item \textbf{Training Execution 3}.  Figure \ref{fig:example_training} (d)
    \begin{itemize}
        \item \textbf{Validate for}: \(\theta_0:\{3\pi/4, 7\pi/4\}, \theta_1:\{\pi/4, 5\pi/4\}\).
        \item \textbf{Lowest energy solution}: \(\theta_0 = 3\pi/4\), \(\theta_1 = 5\pi/4\). \textbf{Accuracy}: 65\%. 
        \item \textbf{Best result}: Execution 3
    \end{itemize}

    \item \textbf{Training Execution 4}.  Figure \ref{fig:example_training} (e)
    \begin{itemize}
        \item \textbf{Validate for}: \(\theta_0:\{3\pi/4, 7\pi/4\}, \theta_1:\{3\pi/4, 7\pi/4\}\).
        \item \textbf{Lowest energy solution}: \(\theta_0 = 3\pi/4\), \(\theta_1 = 7\pi/4\). \textbf{Accuracy}: 55\%. 
        \item \textbf{Best result}: Execution 3
    \end{itemize}
\end{itemize}

\underline{\textbf{Training Level 2}, (Executions 5 to 8)}:
\begin{itemize}
    \item \textbf {Angle range}: \(\pi\). (\(2\pi)\) rescaled 50\%)
    \item \textbf{Centroid}: \(\theta_0=3\pi/4\); \(\theta_1=5\pi/4\).
    \item \textbf{Search space}: \(\theta_0:[\pi/4, 5\pi/4]\); \(\theta_1:[3\pi/4, 7\pi/4]\).
    \item \textbf{Partitions ($d$)}: \(\theta_0:\{[\pi/4 - 3\pi/4], [3\pi/4 - 5\pi/4]\}\); \(\theta_1:\{[3\pi/4 - 5\pi/4], [5\pi/4 - 7\pi/4]\}\).
    \item \textbf{Validation points ($w$)}: \(\theta_0:\{3\pi/8, 7\pi/8\}, \{5\pi/8, 9\pi/8\}\);\\ \(\theta_1:\{7\pi/8, 11\pi/8\}, \{9\pi/8, 13\pi/8\}\)
        
    \item \textbf{Training Execution 5}. 
    \begin{itemize}
        \item \textbf{Validate for}: \(\theta_0:\{3\pi/8, 7\pi/8\}, \theta_1:\{7\pi/8, 11\pi/8\}\).
        \item \textbf{Lowest energy solution}: \(\theta_0 = 7\pi/8\), \(\theta_1 = 7\pi/8\). \textbf{Accuracy}: 62\%.  
        \item \textbf{Best result}: Centroid
    \end{itemize}
    \item \textbf{Training Execution 6}..  Figure \ref{fig:example_training} (f) 
    \begin{itemize}
        \item \textbf{Validate for}: \(\theta_0:\{3\pi/8, 7\pi/8\}, \theta_1:\{9\pi/8, 13\pi/8\}\).
        \item \textbf{Lowest energy solution}: \(\theta_0 = 7\pi/8\), \(\theta_1 = 9\pi/8\). \textbf{Accuracy}: 67\%.  
        \item \textbf{Best result}: Execution 6
    \end{itemize}

    \textbf{...}

    \item \textbf{Training Execution 8}
    \begin{itemize}
        \item \textbf{Validate for}: \(\theta_0:\{5\pi/8, 9\pi/8\}, \theta_1:\{9\pi/8, 13\pi/8\}\).
        \item \textbf{Lowest energy solution}: \(\theta_0 = 5\pi/8\), \(\theta_1 = 13\pi/8\). \textbf{Accuracy}: 64\%. 
        \item \textbf{Best result}: Execution 6
    \end{itemize}
\end{itemize}

\underline{\textbf{Training Level 3}, (Executions 9 to 12)}:
\begin{itemize}
    \item \textbf {Angle range}: \(\pi/2\). (\(\pi\) rescaled 50\%)
    \item \textbf{Centroid}: \(\theta_0=7\pi/8\); \(\theta_1=9\pi/8\).
    \item \textbf{Search space}: \(\theta_0:[5\pi/8, 9\pi/8]\); \(\theta_1:[7\pi/8, 11\pi/8]\).
    \item \textbf{Partitions ($d$)}: \(\theta_0:\{[5\pi/8 - 7\pi/8], [7\pi/8 - 9\pi/8]\}\); \(\theta_1:\{[7\pi/8 - 9\pi/8], [9\pi/8 - 11\pi/8]\}\).
    \item \textbf{Validation points ($w$)}: \(\theta_0:\{11\pi/16, 15\pi/16\}, \{13\pi/16, 17\pi/16\}\);\\ \(\theta_1:\{15\pi/16, 19\pi/16\}, \{17\pi/16, 21\pi/16\}\)
        
    \item \textbf{Training Execution 9}.  Figure \ref{fig:example_training} (g) 
    \begin{itemize}
        \item \textbf{Validate for}: \(\theta_0:\{11\pi/16, 15\pi/16\}, \theta_1:\{15\pi/16, 19\pi/16\}\).
        \item \textbf{Lowest energy solution}: \(\theta_0 = 15\pi/16\), \(\theta_1 = 15\pi/16\). \textbf{Accuracy}: 71\%.  
        \item \textbf{Best result}: Execution 9
    \end{itemize}

    \textbf{...}

    \item \textbf{Training Execution 12}
    \begin{itemize}
        \item \textbf{Validate for}: \(\theta_0:\{13\pi/16, 17\pi/16\}, \theta_1:\{17\pi/16, 21\pi/16\}\).
        \item \textbf{Lowest energy solution}: \(\theta_0 = 13\pi/16\), \(\theta_1 = 21\pi/16\). \textbf{Accuracy}: 64\%. 
        \item \textbf{Best result}: Execution 9
    \end{itemize}
\end{itemize}

\clearpage
\newpage
\section{Multiparameter exploration}\label{ap:parameter_exploration}

\paragraph{Baselines}
Across all graphs, the point at (Levels = 1, Points = 1, Partitions = 1) denotes the baseline configuration, which would be equivalent to a classical brute-force method.

\begin{figure}[htbp]
    \centering
    \begin{tikzpicture}
        \begin{groupplot}[
          group style={
            group size=2 by 2,
            horizontal sep=2cm,
            vertical sep=3cm,
          },
          width=7cm,
          height=6cm,
          xlabel=Partitions and Levels,
          xtick={1,2,3,4},
          ylabel style={align=center}
        ]
    
        % Plot 1: Parts and Levels vs taccuracy
        \nextgroupplot[
          ylabel={Accuracy},
          legend style={font=\small, at={(1.2,-0.3)}, anchor=north, legend columns=2}     
        ]
        % 2 points
        \addplot+[mark=*, color=blue] coordinates {
          (1, 0.6505)
          (2, 0.687)
          (3, 0.675)
          (4, 0.6875)
        }; 
        \addlegendentry{2 points}
        % 4 points
        \addplot+[mark=*, color=red] coordinates {
          (1, 0.65)
          (2, 0.7)
          (3, 0.7)
          (4, 0.7125)
        };
        \addlegendentry{3 points}
        
        % Plot 1.1: Parts and Levels vs taccuracy/time
        \nextgroupplot[
          ylabel={Training Accuracy/Time}
        ]
        % 2 points
        \addplot+[mark=*, color=blue] coordinates {
          (1, 0.000526)
          (2, 0.003251)
          (3, 0.002107)
          (4, 0.001599)
        }; 
        %\addlegendentry{2 points}
        % 3 points
        \addplot+[mark=*, color=red] coordinates {
          (1, 0.000598)
          (2, 0.001120)
          (3, 0.000958)
          (4, 0.000851)
        };
        %\addlegendentry{3 points}
    \end{groupplot}
    \end{tikzpicture}
    \caption{Accuracy and Performance (\textit{accuracy/time}) based on number of training levels and partitions for \textit{Diabetes} dataset with 2 and 3 validation points.}
    \label{fig:diabetes_performance}
\end{figure}
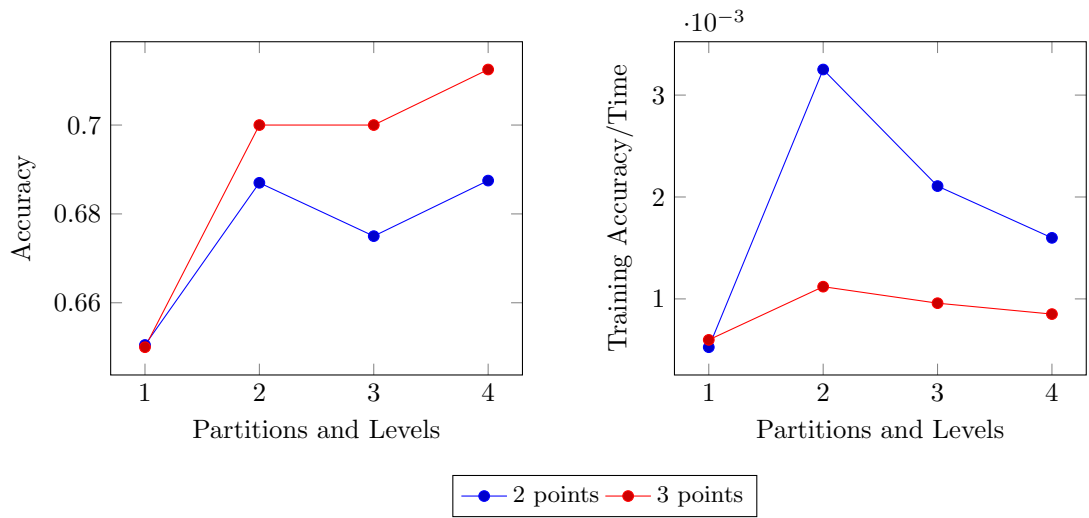

\paragraph{Performance Efficiency} 
The performance efficiency is defined as the amount of accuracy gained per unit of training time, acting as an indicator of the model’s efficiency. Figure~\ref{fig:diabetes_performance} illustrates a general trend of increasing accuracy as training levels and partitions grow. However, higher performance efficiency is observed at lower levels and partitions. Notably, when using 2 validation points, the model achieves both higher accuracy and better performance efficiency compared to the case with 3 validation points.

\subsection{Datasets training results comparison}

\begin{figure}[htbp]
    \centering
    \begin{tikzpicture}
        \begin{groupplot}[
          group style={
            group size=2 by 1,
            horizontal sep=2cm,
          },
          width=6cm,
          height=5cm,
          xlabel=Levels,
          xtick={1,2,3,4},
          ylabel style={align=center},
        ]
        
        % Plot 1: Accuracy
        \nextgroupplot[
          ylabel={Accuracy},
          legend style={font=\small, at={(1.2,-0.3)}, anchor=north, legend columns=3}
        ]
        \addplot+[mark=*, color=blue] coordinates {
          (1, 0.7344) (2, 0.7187) (3, 0.7656) (4, 0.7969)
        };
        \addlegendentry{Iris}
        
        \addplot+[mark=*, color=red] coordinates {
          (1, 0.7125) (2, 0.7) (3, 0.7) (4, 0.7125)
        };
        \addlegendentry{Diabetes}
        
        \addplot+[mark=*, color=green] coordinates {
          (1, 0.639) (2, 0.64875) (3, 0.64725) (4, 0.655)
        };
        \addlegendentry{Heart Disease}
        
        % Plot 2: Time
        \nextgroupplot[
          ylabel={Time},
        ]
        \addplot+[mark=*, color=blue] coordinates {
          (1, 503.54) (2, 506.13) (3, 427.36) (4, 191.38)
        };
        
        \addplot+[mark=*, color=red] coordinates {
          (1, 531.52) (2, 638.39) (3, 730.28) (4, 836.91)
        };
        
        \addplot+[mark=*, color=green] coordinates {
          (1, 1769.6725) (2, 1762.04) (3, 1762.755) (4, 1746.08325)
        };
        \end{groupplot}
    \end{tikzpicture}
    \caption{Accuracy and Cost (Time) based on Training Levels. Partitions and validation points are fixed to 3.}
    \label{fig:levels_accuracy_time}
\end{figure}
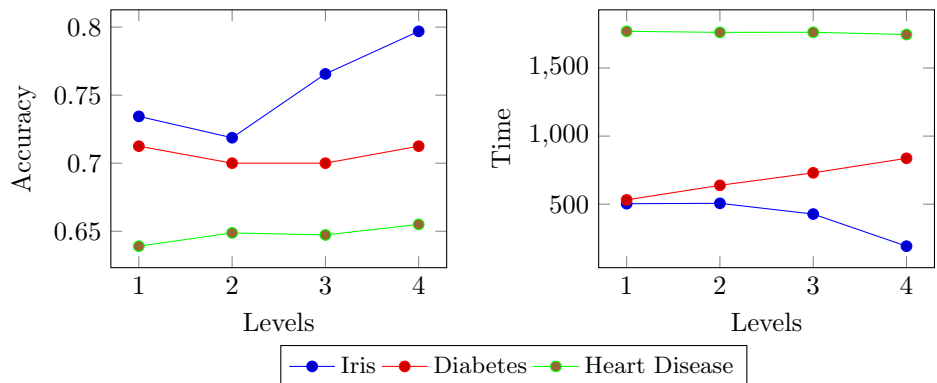

\subsection{\textit{Diabetes} dataset details}

\begin{figure}[htbp]
    \centering
    \begin{tikzpicture}
    \begin{groupplot}[
      group style={
        group size=2 by 1,
        horizontal sep=2cm,
      },
      width=6cm,
      height=5cm,
      xlabel=Validation Points,
      xtick={1,2,3,4,5,6},
      ylabel style={align=center},
    ]
    
    % Plot 1: Points vs taccuracy 
    \nextgroupplot[
      ylabel={Accuracy},
      legend style={font=\small, at={(1.2,-0.3)}, anchor=north, legend columns=3}
    ]
    \addplot+[mark=*, color=blue] coordinates {
      (1, 0.65)
      (2, 0.6975)
      (3, 0.7125)
      (4, 0.72625)
      (5, 0.7333)
    };
    \addlegendentry{Diabetes}
    \addplot+[mark=*, color=orange] coordinates {
      (1, 0.937)
      (2, 0.906)
      (3, 0.890)
      (4, 0.921)
      (5, 0.921)
    };
    \addlegendentry{Iris}
    \addplot+[mark=*, color=green] coordinates {
      (1, 0.671)
      (2, 0.6645)
      (3, 0.669)
      (4, 0.670675)
      (5, 0.6815)
    };
    \addlegendentry{Heart Disease}
    
    % Plot 2: Points vs Time
    \nextgroupplot[
      ylabel={Time},
    ]
    \addplot+[mark=*, color=blue] coordinates {
      (1, 6.9312)
      (2, 106.901)
      (3, 531.52)
      (4, 1644.08)
      (5, 4107.24)
    };
    %\addlegendentry{Diabetes}
    \addplot+[mark=*, color=orange] coordinates {
      (1, 115.67)
      (2, 114.03)
      (3, 115.795)
      (4, 356.8525)
      (5, 874.6)
    };
    %\addlegendentry{Iris}
    \addplot+[mark=*, color=green] coordinates {
      (1, 17261.2275)
      (2, 17484.7725)
      (3, 17324.53)
      (4, 17001.315)
      (5, 17333.58)
    };
    %\addlegendentry{Heart Disease}
    \end{groupplot}
    \end{tikzpicture}
    \caption{Accuracy and Cost (Time) based on Validation points for \textit{Diabetes} dataset.  Levels are fixed to 1 and Partitions are fixed to 2.}
    \label{fig:points_accuracy_time}
\end{figure}
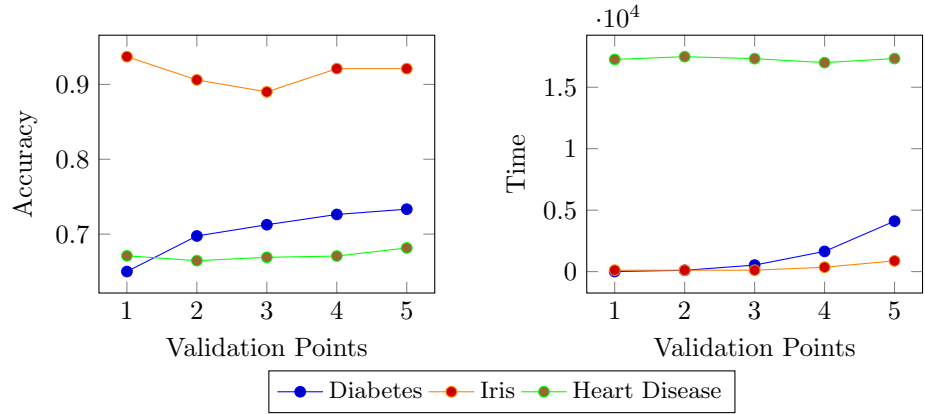

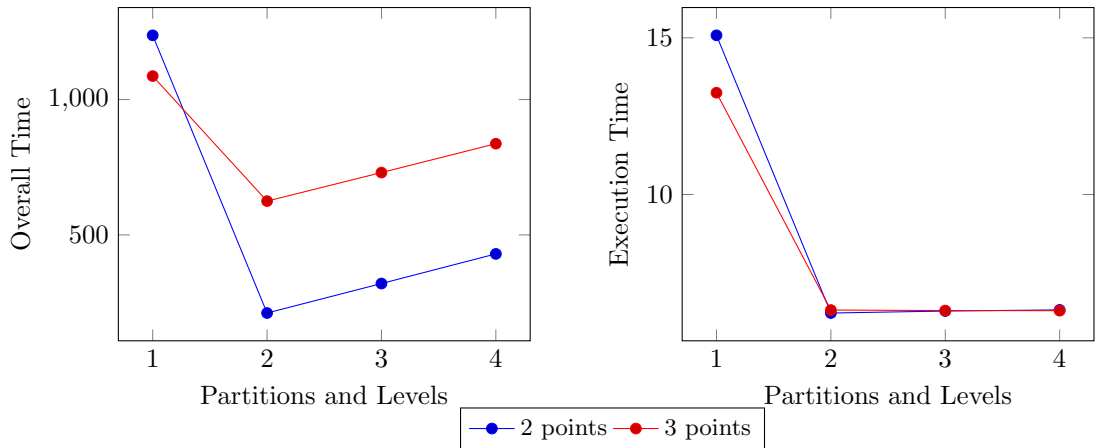
\begin{figure}[htbp]
    \centering
    \begin{tikzpicture}
        \begin{groupplot}[
          group style={
            group size=2 by 2,
            horizontal sep=2cm,
            vertical sep=3cm,
          },
          width=7cm,
          height=6cm,
          xlabel=Partitions and Levels,
          xtick={1,2,3,4},
          ylabel style={align=center}
        ]
        % Plot 2: Overall Time
        \nextgroupplot[
          ylabel={Overall Time},
          legend style={font=\small, at={(1.2,-0.2)}, anchor=north, legend columns=2},
        ]
        % 2 points
        \addplot+[mark=*, color=blue] coordinates {
          (1, 1237.296)
          (2, 211.38)
          (3, 320.37)
          (4, 429.75)
        };
        \addlegendentry{2 points}
        % 3 points
        \addplot+[mark=*, color=red] coordinates {
          (1, 1086.728)
          (2, 624.848)
          (3, 730.28)
          (4, 836.91)
        };
        \addlegendentry{3 points}
        
        % Plot 3: Partitions vs Time
        \nextgroupplot[
          ylabel={Execution Time}
        ]
        % 1 Level, 2 points
        \addplot+[mark=*, color=blue] coordinates {
          (1, 15.08)
          (2, 6.21706)
          (3, 6.28078)
          (4, 6.31765)
        };
        %\addlegendentry{2 points}
        % 3 points
        \addplot+[mark=*, color=red] coordinates {
          (1, 13.248)
          (2, 6.3126)
          (3, 6.29638)
          (4, 6.29173)
        };
        %\addlegendentry{3 points}
        \end{groupplot}
    \end{tikzpicture}
    \caption{Total training time and training time per execition based on number of training levels and partitions for \textit{Diabetes} dataset with 2 and 3 validation points.}
    \label{fig:diabetes_training_times}
\end{figure}

\begin{figure}[htbp]
    \centering
    \begin{tikzpicture}
    \begin{groupplot}[
      group style={
        group size=2 by 1,
        horizontal sep=2cm,
      },
      width=7cm,
      height=6cm,
      xlabel=Levels,
      xtick={1,2,3,4},
      ylabel style={align=center}
    ]
    
    % Plot 1: Partitions vs taccuracy
    \nextgroupplot[
      ylabel={Accuracy},
      legend style={font=\small, at={(1.2,-0.2)}, anchor=north, legend columns=4},
    ]
    \addplot+[mark=*, color=blue] coordinates {
      (1, 0.7125)
      (2, 0.7)
      (3, 0.7)
      (4, 0.7125)
    };
    \addlegendentry{3 parts, 3 points}
    \addplot+[mark=*, color=red] coordinates {
      (1, 0.65)
      (2, 0.6375)
      (3, 0.658)
      (4, 0.675)
    };
    \addlegendentry{2 parts, 1 point}
    \addplot+[mark=*, color=green] coordinates {
      (1, 0.6975)
      (2, 0.687)
      (3, 0.6875)
      (4, 0.7)
    };
    \addlegendentry{2 parts, 2 points}
    \addplot+[mark=*, color=orange] coordinates {
      (1, 0.7125)
      (2, 0.7)
      (3, 0.7)
      (4, 0.7)
    };
    \addlegendentry{2 parts, 3 points}
    \addplot+[mark=*, color=brown] coordinates {
      (1, 0.6875)
      (2, 0.685)
      (3, 0.675)
      (4, 0.6875)
    };
    \addlegendentry{3 parts, 2 points}
    \addplot+[mark=*, color=cyan] coordinates {
      (1, 0.6625)
      (2, 0.6608)
      (3, 0.668)
      (4, 0.656)
    };
    \addlegendentry{4 parts, 3 points}
    \addplot+[mark=*, color=yellow] coordinates {
      (1, 0.671)
      (2, 0.668)
      (3, 0.6625)
      (4, 0.6675)
    };
    \addlegendentry{5 parts, 3 points}
    
    % Plot 2: Parts vs Time
    \nextgroupplot[
      ylabel={Time},
    ]
    \addplot+[mark=*, color=blue] coordinates {
      (1, 514.866)
      (2, 638.39)
      (3, 730.28)
      (4, 836.91)
    };
    %\addlegendentry{3 parts, 3 points}
    \addplot+[mark=*, color=red] coordinates {
      (1, 6.9312)
      (2, 14.15)
      (3, 21.51)
      (4, 28.77)
    };
    %\addlegendentry{2 parts, 1 point}
    \addplot+[mark=*, color=green] coordinates {
      (1, 106.901)
      (2, 211.38)
      (3, 319.33)
      (4, 411.75)
    };
    %\addlegendentry{2 parts, 2 points}
    \addplot+[mark=*, color=green] coordinates {
      (1, 531.52)
      (2, 624.848)
      (3, 714.93)
      (4, 840.69)
    };
    %\addlegendentry{2 parts, 3 points}
    \addplot+[mark=*, color=brown] coordinates {
      (1, 104.454)
      (2, 209.67)
      (3, 320.37)
      (4, 429.75)
    };
    %\addlegendentry{3 parts, 2 points}
    \addplot+[mark=*, color=cyan] coordinates {
      (1, 710.55)
      (2, 703.45)
      (3, 796.50)
      (4, 2024.63)
    };
    %\addlegendentry{4 parts, 3 points}
    \addplot+[mark=*, color=yellow] coordinates {
      (1, 1223.62)
      (2, 1482.18)
      (3, 1740.9)
      (4, 2004.11)
    };
    %\addlegendentry{5 parts, 3 points}
    
    \end{groupplot}
    \end{tikzpicture}
    \caption{Training accuracy and time based for different training levels for \textit{Diabetes}.}
    \label{fig:diabetes_training_times}
\end{figure}
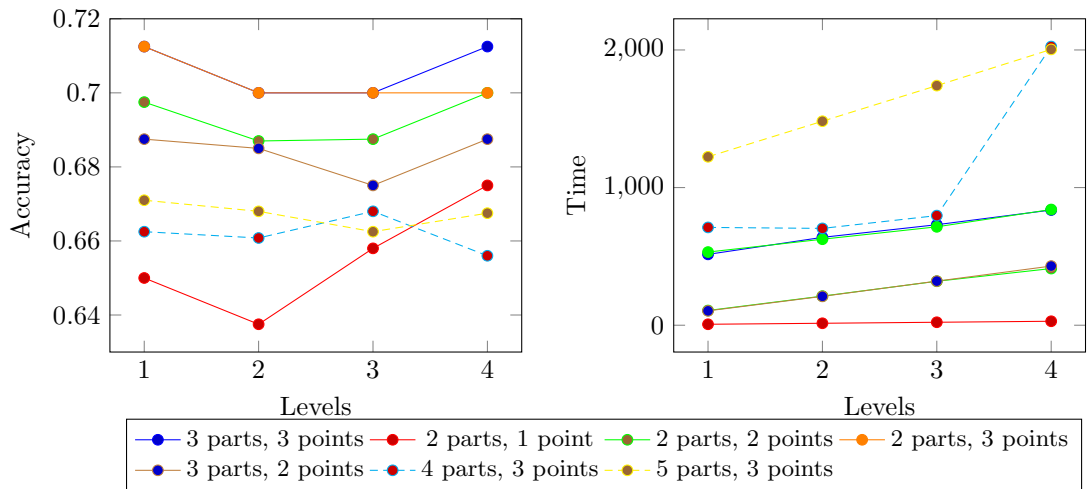

\clearpage
\section{Data results}\label{ap:parameter_exploration_data}

\subsection{Multiparameter exploration data results} 

\begin{table}[htbp]
\centering
\caption{Averaged \textit{Iris} Results}
\begin{filecontents*}{csv/iris_results_detailed_avgs.csv}
label,dataset,size,levels,parts,points,taccuracy,time,iterations,vaccuracy,texec,acctime
A,iris,60,4,2,2,0.7708,93.783,68,0.75,1.3792,0.00822
B,iris,60,3,3,3,0.625,161.553,116,0.6,1.3927,0.00387
C,iris,100,2,3,4,0.8125,411.98,264,0.5167,1.5605,0.00197
D,iris,100,1,4,5,0.6875,1203.11,626,0.5667,1.9226,0.00057
E,iris,100,1,3,3,0.7344,503.54,133,0.5625,3.7853,0.00146
F,iris,400,2,3,3,0.7187,506.13,133,0.525,3.805,0.00142
G,iris,400,3,3,3,0.7656,427.36,133,0.5,3.21,0.00179
H,iris,400,4,3,3,0.7969,191.38,133,0.5,1.43,0.00416
I,iris,100,1,2,1,0.937,115.67,82,0.6,1.41,0.00810
J,iris,100,1,2,2,0.906,114.03,82,0.5,1.39,0.00794
K,iris,100,1,2,3,0.890625,115.795,82,0.5125,1.4115,0.00769
L,iris,100,1,2,4,0.921875,356.8525,257,0.4875,1.3882,0.00258
M,iris,100,1,2,5,0.921875,874.6,626,0.5125,1.3965,0.00105
N,iris,100,1,2,6,0.68325,1796.7225,1297,0.6225,13.8466,0.00038
O,iris,100,1,1,1,0.4625,2.05,2,0.54,2.05,0.22561
\end{filecontents*}
\pgfplotstabletypeset[
    col sep=comma,
    string type,
    columns={label,levels,parts,points,taccuracy,time,iterations,texec,vaccuracy,acctime},
    columns/size/.style={column name=Label},
    columns/levels/.style={column name=Levels},
    columns/parts/.style={column name=Parts},
    columns/points/.style={column name=Points},
    columns/taccuracy/.style={column name=\shortstack{Train\\Accuracy}},
    columns/time/.style={column name=\shortstack{Training\\Time}},
    columns/iterations/.style={column name=Executions},
    columns/texec/.style={column name=\shortstack{Time per\\ Exec}},
    columns/acctime/.style={column name=\shortstack{Accuracy/\\Time}},
    columns/vaccuracy/.style={column name=\shortstack{Validation\\Accuracy}},
    every head row/.style={before row=\toprule, after row=\midrule},
    every last row/.style={after row=\bottomrule},
]{csv/iris_results_detailed_avgs.csv}
\end{table}

\begin{table}[htbp]
\centering
\caption{Averaged \textit{Heart Disease} Results}
\begin{filecontents*}{csv/heartdisease_results_detailed_avgs.csv}
label,dataset,size,levels,parts,points,taccuracy,time,iterations,vaccuracy,texec,acctime
A,heart,400,4,2,2,0.6463,373.7267,68.0,0.59,5.49657,0.00173
B,heart,400,3,3,3,0.6760,630.3367,116.0,0.6167,5.43324,0.00107
C,heart,1000,2,3,4,0.6263,3669.86,339.0,0.5540,10.8236,0.00017
D,heart,1000,1,4,5,0.6063,7125.53,626.0,0.5920,11.3781,0.00009
E,heart,1000,1,3,3,0.639,1769.6725,133,0.64125,13.2983,0.00036
F,heart,1000,2,3,3,0.64875,1762.04,133,0.58875,13.2436,0.00037
G,heart,1000,3,3,3,0.64725,1762.755,133,0.6275,13.2479,0.00037
H,heart,1000,4,3,3,0.655,1746.08325,133,0.6475,13.1232,0.00037
I,heart,1000,1,2,1,0.671,17261.2275,1297,0.63,13.3076,0.00004
J,heart,1000,1,2,2,0.6645,17484.7725,1297,0.64225,13.4789,0.00004
K,heart,1000,1,2,3,0.669,17324.53,1297,0.64475,13.3561,0.00004
L,heart,1000,1,2,4,0.670675,17001.315,1297,0.64225,13.1059,0.00004
M,heart,1000,1,2,5,0.6815,17333.58,1297,0.6375,13.3637,0.00004
N,heart,1000,1,2,6,0.6815,16787.15,1297,0.645,12.9408,0.00004
O,heart,400,1,1,1,0.5298,14.18,2.0,0.518,14.18,0.03734
\end{filecontents*}
\pgfplotstabletypeset[
    col sep=comma,
    string type,
    columns={label,levels,parts,points,taccuracy,time,iterations,texec,vaccuracy,acctime},
    columns/size/.style={column name=Label},
    columns/levels/.style={column name=Levels},
    columns/parts/.style={column name=Parts},
    columns/points/.style={column name=Points},
    columns/taccuracy/.style={column name=\shortstack{Train\\Accuracy}},
    columns/time/.style={column name=\shortstack{Training\\Time}},
    columns/iterations/.style={column name=Executions},
    columns/texec/.style={column name=\shortstack{Time per\\ Exec}},
    columns/acctime/.style={column name=\shortstack{Accuracy/\\Time}},
    columns/vaccuracy/.style={column name=\shortstack{Validation\\Accuracy}},
    every head row/.style={before row=\toprule, after row=\midrule},
    every last row/.style={after row=\bottomrule},
]{csv/heartdisease_results_detailed_avgs.csv}
\end{table}

\begin{table}[htbp]
\centering
\caption{Averaged \textit{Diabetes} Results} 
\begin{filecontents*}{csv/diabetes_results_detailed_avgs.csv}
label,dataset,size,levels,parts,points,taccuracy,time,iterations,vaccuracy,texec,acctime
A,diabetes,400,1,2,1,0.65,6.9312,2,0.6945,3.4656,0.093775
B,diabetes,400,1,2,2,0.6975,106.901,17,0.6472,6.28829,0.006525
B1,diabetes,400,1,3,2,0.6875,104.454,17,0.652,6.14435,0.006582
B2,diabetes,400,1,4,2,0.6375,115.55,17,0.62,6.79706,0.005516
B3,diabetes,400,1,5,2,0.6525,242.606,17,0.606,14.271,0.002689
B4,diabetes,400,1,6,2,0.6025,2430.09,17,0.538,142.94647,0.000248
C,diabetes,400,1,2,3,0.7125,531.52,82,0.59,6.48293,0.001340
D,diabetes,400,1,2,4,0.72625,1644.08,228.57,0.626,7.19338,0.000442
D1,diabetes,400,1,3,4,0.7175,1658.16,257,0.636,6.45238,0.000433
D2,diabetes,400,1,4,4,0.675,1848.12,257,0.558,7.19112,0.000365
D3,diabetes,400,1,5,4,0.675,3815.33,257,0.58,14.84762,0.000177
E,diabetes,400,1,2,5,0.7333,4107.24,626,0.6678,6.56246,0.000179
E1,diabetes,400,1,2,6,0.729,8264.75,1297,0.63,6.37063,0.000088
F,diabetes,400,1,3,1,0.625,7.008,2,0.62,3.504,0.089204
G,diabetes,400,1,3,2,0.678,107.35,17,0.63,6.31471,0.006314
H,diabetes,400,1,3,3,0.7125,514.866,82,0.63,6.27873,0.001384
I,diabetes,400,2,2,1,0.6375,14.15,4,0.71,3.5375,0.045029
J,diabetes,400,2,2,2,0.687,211.38,34,0.55,6.21706,0.003251
K,diabetes,400,2,2,3,0.7,624.848,99,0.67,6.3126,0.001120
L,diabetes,400,2,3,1,0.6125,14.253,4,0.64,3.56325,0.042983
M,diabetes,400,2,3,2,0.685,209.67,34,0.618,6.16735,0.003267
N,diabetes,400,2,3,3,0.7,638.39,99,0.69,6.44838,0.001096
Z3,diabetes,400,3,2,1,0.658,21.51,6,0.689,3.585,0.030597
P,diabetes,400,3,2,2,0.6875,319.33,51,0.63,6.25941,0.002153
Q,diabetes,400,3,2,3,0.7,714.93,116,0.62,6.16233,0.000979
R,diabetes,400,3,3,1,0.625,21.52,6,0.63,3.58667,0.029048
S,diabetes,400,3,3,2,0.675,320.37,51,0.64,6.28078,0.002107
T,diabetes,400,3,3,3,0.7,730.28,116,0.61,6.29638,0.000958
U,diabetes,400,4,2,1,0.675,28.77,8,0.69,3.59625,0.023456
V,diabetes,400,4,2,2,0.7,411.75,68,0.68,6.05441,0.001700
W,diabetes,400,4,2,3,0.7,840.69,133,0.68,6.32022,0.000833
X,diabetes,400,4,3,1,0.625,28.601,8,0.61,3.57513,0.021859
Y,diabetes,400,4,3,2,0.6875,429.75,68,0.61,6.31765,0.001599
Z,diabetes,400,4,3,3,0.7125,836.91,133,0.59,6.29173,0.000851
Z11,diabetes,400,1,4,3,0.6625,710.55,99,0.6125,7.17,0.000932
Z12,diabetes,400,2,4,3,0.6608,703.45,99,0.6125,7.10,0.000939
Z13,diabetes,400,3,4,3,0.66875,796.50,116,0.58,6.86,0.000840
Z1,diabetes,400,4,4,3,0.655,2024.63,133,0.592,15.22,0.000323
Z21,diabetes,400,1,5,3,0.671,1223.62,82,0.587,14.92,0.000548
Z22,diabetes,400,2,5,3,0.668,1482.18,99,0.605,14.97,0.000451
Z23,diabetes,400,3,5,3,0.6625,1740.9,166,0.625,15.00,0.000381
Z2,diabetes,400,4,5,3,0.6675,2004.11,133,0.606,15.06,0.000333
O,diabetes,400,1,1,1,0.5105,14.122,2.0,0.504,14.122,0.036144
Z4,diabetes,400,1,1,2,0.6505,1237.296,82.0,0.66,15.08,0.000526
Z6,diabetes,400,1,1,3,0.65,1086.728,82.0,0.628,13.248,0.000598
Z5,diabetes,400,1,1,4,0.638,1072.736,82.0,0.647,13.078,0.000594
\end{filecontents*}
\pgfplotstabletypeset[
    col sep=comma,
    string type,
    columns={label,levels,parts,points,taccuracy,time,iterations,texec,vaccuracy,acctime},
    columns/label/.style={column name=Label},
    columns/levels/.style={column name=Levels},
    columns/parts/.style={column name=Parts},
    columns/points/.style={column name=Points},
    columns/taccuracy/.style={column name=\shortstack{Training\\Accuracy}},
    columns/time/.style={column name=\shortstack{Training\\Time}},
    columns/iterations/.style={column name=Executions},
    columns/texec/.style={column name=\shortstack{Time per\\Exec}},
    columns/acctime/.style={column name=\shortstack{Accuracy/\\Time}},
    columns/vaccuracy/.style={column name=\shortstack{Validation\\Accuracy}},
    every head row/.style={
        before row=\toprule,
        after row=\midrule,
    },
    every last row/.style={after row=\bottomrule},
]{csv/diabetes_results_detailed_avgs.csv}
\end{table}

\clearpage
\newpage
\subsection{Training experiments}\label{ap:training_results}

\begin{table}[htbp]
\centering
\caption{Adiabatic training on \textit{Iris} dataset}
\begin{tabular}{llllllllll}
\toprule
Angle & Measured & Ansatz & Parts & Search & Dataset & Accuracy & Time(s) & Total & t/iter/ \\
range & qubit & (gates) & & Deep & Length & & & Iterations & record(s) \\
\midrule
$-2\pi, 2\pi$ & All & 4 & 2 & 4 & 80 & 0.52 & 20.04 & 3.2 & 0.078 \\
$-6\pi, 6\pi$ & All & 4 & 2 & 4 & 80 & \textbf{0.59} & 23.12 & 3.6 & 0.080 \\
$-2\pi, 2\pi$ & 2 & 4 & 2 & 4 & 80 & 0.518 & 24.96 & 3.9 & 0.08 \\
$-2\pi, 2\pi$ & 2 & 4 & 2 & 5 & 80 & 0.56 & 31.87 & 5 & 0.079 \\
$-2\pi, 2\pi$ & 2 & 4 & 2 & 4 & 52 & 0.58 & \textbf{15.81} & 3 & 0.101 \\
$-2\pi, 2\pi$ & 1 & 4 & 2 & 4 & 80 & 0.48 & 23.42 & 3.7 & 0.079 \\
$-2\pi, 2\pi$ & All & 3 & 2 & 4 & 80 & 0.462 & 17.62 & 3.5 & \textbf{0.063} \\
$-2\pi, 2\pi$ & 2 & 4 & 3 & 4 & 80 & 0.475 & 135.16 & 3.6 & 4.693 \\
$-2\pi, 2\pi$ & All & 4 & 3 & 4 & 80 & 0.55 & 137.52 & 3.8 & 0.452 \\
\bottomrule
\end{tabular}
\label{tab:iris_adiabatic_results}
\end{table}

\begin{table}[htbp]
\centering
\caption{Adiabatic training on \textit{Heart Disease} dataset}
\begin{tabular}{cccccccccc}
\toprule
\shortstack{Angle\\range} & \shortstack{Measured\\qubit} & \shortstack{Ansatz\\(gates)} & Parts & \shortstack{Search\\Deep} & \shortstack{Dataset\\Length} & Accuracy & \shortstack{Time\\(s)} & Iterations & \shortstack{t/iter\\record(s)} \\
\midrule
$-2\pi,2\pi$ & All & 4 & 2 & 4 & 106 & 0.6495 & 445.94 & 4 & 1.051 \\
$-6\pi,6\pi$ & All & 4 & 2 & 4 & 106 & 0.6295 & 407.55 & 3.8 & 1.012 \\
$-2\pi,2\pi$ & 2 & 4 & 2 & 4 & 106 & \textbf{0.7568} & 438.09 & 4 & 1.033 \\
$-2\pi,2\pi$ & 2 & 4 & 2 & 5 & 100 & 0.6954 & 548.28 & 5 & 1.034 \\
$-2\pi,2\pi$ & 2 & 4 & 2 & 4 & 1052 & 0.6516 & 4177.43 & 4 & 0.993 \\
$-2\pi,2\pi$ & 1 & 4 & 2 & 4 & 106 & 0.54 & 498.91 & 3.6 & 1.307 \\
$-2\pi,2\pi$ & All & 3 & 2 & 4 & 106 & 0.6136 & 191.72 & 4 & \textbf{0.452} \\
$-2\pi,2\pi$ & 2 & 4 & 3 & 4 & 106 & 0.74 & 1752.26 & 4 & 4.133 \\
$-2\pi,2\pi$ & All & 4 & 3 & 4 & 106 & 0.6636 & 1650.11 & 3.8 & 3.891 \\
\bottomrule
\end{tabular}
\label{tab:heart_adiabatic_results}
\end{table}

\begin{table}[htbp]
\centering
\caption{Adiabatic training on \textit{Diabetes} dataset}
\begin{tabular}{cccccccccc}
\toprule
\shortstack{Angle\\range} & \shortstack{Measured\\qubit} & \shortstack{Ansatz\\(gates)} & Parts & \shortstack{Search\\Deep} & \shortstack{Dataset\\Length} & Accuracy & \shortstack{Time\\(s)} & Iterations & \shortstack{t/iter\\record(s)} \\
\midrule
$-2\pi,2\pi$ & All & 4 & 2 & 4 & 108 & 0.519 & 16.27 & 2.2 & 0.068 \\
$-6\pi,6\pi$ & All & 4 & 2 & 4 & 108 & 0.718 & 18.95 & 3 & 0.058 \\
$-2\pi,2\pi$ & 2 & 4 & 2 & 4 & 102 & 0.603 & 30.46 & 4 & 0.074 \\
$-2\pi,2\pi$ & 2 & 4 & 2 & 5 & 97 & \textbf{0.725} & 36.22 & 5 & 0.074 \\
$-2\pi,2\pi$ & All & 4 & 2 & 4 & 800 & 0.452 & 138.85 & 4 & 0.043 \\
$-2\pi,2\pi$ & 1 & 4 & 2 & 4 & 107 & 0.709 & 30.92 & 4 & 0.072 \\
$-2\pi,2\pi$ & All & 3 & 2 & 4 & 99 & 0.564 & 14.69 & 3 & 0.049 \\
$-2\pi,2\pi$ & 2 & 4 & 3 & 4 & 112 & 0.685 & 144.69 & 4 & 0.323 \\
$-2\pi,2\pi$ & All & 4 & 3 & 4 & 102 & 0.547 & 138.14 & 3.9 & \textbf{0.034} \\
\bottomrule
\end{tabular}
\label{tab:diabetes_adiabatic_results}
\end{table}

\clearpage
\newpage

\FloatBarrier




\end{appendices}

%\clearpage
%\bibliography{qaa4vqa-bibliography}% common bib file

\end{document}